\documentclass[pre,reprint,showpacs,superscriptaddress,floatfix]{revtex4-1}
\usepackage{amsmath}
\usepackage{amssymb}
\usepackage{amstext}
\usepackage{color}
\usepackage{graphicx}
\usepackage{adjustbox}

\newcommand{\um}[1]{\;\mathrm{#1}}
\newcommand{\ket}[1]{\left\vert #1\,\right\rangle}

\newcommand{\bra}[1]{\left\langle #1\,\right\vert}

\begin{document}
\title[Optimal efficiency of quantum transport in a disordered trimer]{Optimal efficiency of quantum transport in a disordered trimer}

% %%%%Infos for APS journals
 \date{\today}
 \author{Giulio~G. \surname{Giusteri}}
 \email{giulio.giusteri@oist.jp}
 \affiliation{Mathematical Soft Matter Unit, Okinawa Institute of Science and Technology Graduate University, 1919-1 Tancha, Onna-son, Kunigami-gun, Okinawa, Japan 904-0495}
 \affiliation{Dipartimento di Matematica e Fisica and Interdisciplinary Laboratories for Advanced Materials Physics, Universit\`a Cattolica del Sacro Cuore, via Musei 41, I-25121 Brescia, Italy \\ and Istituto Nazionale di Fisica Nucleare, Sezione di Pavia, via Bassi 6, I-27100,  Pavia, Italy}
 % \affiliation{International Research Center on Mathematics \& Mechanics of Complex Systems, via XIX marzo 1 I-04012 Cisterna di Latina, Italy}
 \author{G.~Luca \surname{Celardo}}
 \affiliation{Dipartimento di Matematica e Fisica and Interdisciplinary Laboratories for Advanced Materials Physics, Universit\`a Cattolica del Sacro Cuore, via Musei 41, I-25121 Brescia, Italy \\ and Istituto Nazionale di Fisica Nucleare, Sezione di Pavia, via Bassi 6, I-27100,  Pavia, Italy}
 \author{Fausto \surname{Borgonovi}}
 \affiliation{Dipartimento di Matematica e Fisica and Interdisciplinary Laboratories for Advanced Materials Physics, Universit\`a Cattolica del Sacro Cuore, via Musei 41, I-25121 Brescia, Italy \\ and Istituto Nazionale di Fisica Nucleare, Sezione di Pavia, via Bassi 6, I-27100,  Pavia, Italy}

%%%%%Infos for IOP journals

%\author{G G Giusteri$^{1,2}$, G L Celardo$^{2}$ and F Borgonovi$^{2}$}
%\address{$^1$Mathematical Soft Matter Unit, Okinawa Institute of Science and Technology Graduate University, 1919-1 Tancha, Onna, 904-0495, Okinawa, Japan}
%\address{$^2$Dipartimento di Matematica e Fisica and Interdisciplinary Laboratories for Advanced Materials Physics, Universit\`a Cattolica del Sacro Cuore, via Musei 41, I-25121 Brescia, Italy 
%\emph{and}
%Istituto Nazionale di Fisica Nucleare, Sezione di Pavia, via Bassi 6, I-27100,  Pavia, Italy}
%\ead{giulio.giusteri@oist.jp}

\begin{abstract}
Disordered quantum networks, as those describing 
light-harvesting complexes, are often characterized 
by the presence of peripheral ring-like structures, where the excitation 
is initialized, and inner structures, reaction centers (RC), where
the excitation is trapped and transferred. The peripheral rings often display distinguished
coherent features: 
their eigenstates  can be separated, with respect to the transfer of 
excitation, in the two classes of superradiant and subradiant states. 
Both are important to optimize transfer efficiency.
In the absence of disorder, superradiant states have an enhanced coupling strength to the
RC, while the subradiant ones are basically decoupled from it. 
Static on-site disorder induces a coupling between subradiant and superradiant
states, thus creating an indirect coupling to the RC. 
The problem of finding the optimal transfer
conditions, as a function of both the RC energy and the disorder strength,
is very complex even in the simplest network, namely a three-level system.
In this paper we analyze such trimeric structure choosing as initial condition
an excitation on a subradiant state, rather than the more common choice
of an excitation localized on a single site.
We show that, while the optimal disorder is of
the order of the superradiant coupling, the optimal
detuning between the initial state and the RC energy strongly depends
on system parameters: when the superradiant coupling is much larger
than the energy gap between
the superradiant and the subradiant levels, optimal transfer occurs
if the RC energy is at resonance with the subradiant initial state,
whereas we find
an optimal RC energy at resonance with a virtual dressed state when the superradiant coupling is smaller than or comparable with the gap.
The presence of dynamical noise, which 
induces dephasing and decoherence, affects the resonance structure
of energy transfer producing an additional ``incoherent'' resonance peak, which corresponds
to the RC energy being equal to the energy of the superradiant state.
\end{abstract}

%\noindent{\it Keywords\/}: quantum transport in disordered systems;
%open quantum systems; energy transfer optimization. 

\maketitle
%\ioptwocol

\section{Introduction}

Photosynthetic bacteria utilize antenna complexes to capture photons
and convert the energy of the short-lived electronic excitation in a more stable form,
such as chemical bonds. After absorption, the energy is
transferred to a complex, called reaction center (RC), where 
it initiates electron transfer, resulting in a
membrane potential. 
%This energy migration process has been 
%described by F\"orster \cite{forster1, forster2}
%and it is called Fluorescence Resonance Energy Transfer.
This very efficient transfer occurs on a time-scale of few hundreds of picoseconds and on a length-scale of few nanometers, so that  
coherent quantum dynamics can enter the play, as recent experiments seem to
prove~\cite{effnat}. 
Quantum coherence can
enhance transport efficiency inducing Supertransfer and Superradiance in light-harvesting complexes~\cite{srfmo,srrc,others}.
On the other hand, quantum coherence can also be
detrimental to transport, as Anderson localization~\cite{Anderson} and the presence of trapping-free subspaces~\cite{invsub} show.

Superradiance~\cite{Zannals,rottertb,puebla}, as viewed in
the context of both optical
fluorescence \cite{mukameldeph,robin,vangrondelle} and quantum
transport in open systems \cite{kaplan,srfmo,alberto,prbdisorder,prbdephasing}, is not solely a many-body effect. 
Single-excitation superradiance
is a prominent example of genuinely quantum cooperative
effect~\cite{scully}, relevant in natural complexes,
which operate in the single excitation regime since
solar light is very dilute.

Natural complexes are subject  to a noisy environment with different correlation time-scales (if compared to the excitonic transport time): 
(i) short-time correlations, giving rise to dephasing
(homogeneous broadening, as considered e.g.\ 
in \cite{enaqt6, enaqt3, enaqt4, enaqt9})
and (ii) long-time correlations, producing on-site static disorder
(inhomogeneous broadening, as considered e.g.\ in \cite{disorder,fassioli}).
Following a common nomenclature, we will refer to the former effect as 
dephasing noise and to the latter as static disorder.

The role of environment is twofold: on one hand,
it can help transport since it destroys the detrimental coherent
effects, leading to noise-enhanced energy transfer,
i.e.\ the existence of a maximal efficiency at some intermediate
noise strength, as found in the last decade by various groups
\cite{enaqt1,enaqt2,enaqt3,enaqt4,invsub,enaqt5,enaqt6,enaqt7,enaqt8,plenioPTA,enaqt9,caoprl}. On the other hand, it can suppress the beneficial coherences leading to a
quenching of Supertransfer~\cite{prbdisorder}. It is thus essential to
consider this non-trivial interplay.

Typical structures of bacterial photosynthetic complexes display a
RC placed at the center of the light-harvesting complex I (LHI), with the
chromophores arranged on a ring and surrounded by other ring
structures (called LHII) acting as peripheral antennae.
The LHI-RC structure describing light-harvesting complexes produces a distinguished feature:
the eigenstates of the peripheral ring structure can be separated into
two classes, superradiant and subradiant states, with respect to
the transfer of excitation towards the RC. In absence of both environmental noise and static disorder, 
the superradiant states are
coupled to the RC with a coupling amplitude proportional to
$\sqrt{N}$, where $N$ is the number of chromophores in the ring, while
subradiant states are basically decoupled from the RC. 
In presence of disorder, the subradiant states can be coupled to the
superradiant ones and, as a consequence, only indirectly coupled to
the RC states. Note that the subradiant subspace, previously studied
by the authors~\cite{prbdisorder}, has been also analyzed in the
literature under the name of trapping-free subspace~\cite{invsub}. 

The optimization of excitation transfer efficiency from peripheral states in networks 
displaying the Ring-RC structure is a rather difficult problem, even for a network as simple as
a trimer. 
Despite its simplicity, the trimer model has been discussed in several papers as a paradigmatic model \cite{newref,enaqt6}. 
In particular, in~\cite{enaqt6} various optimal conditions
are explored in simple few-site networks in the presence of dephasing noise but without static disorder.
{Since it is difficult to clarify generic physical effects by studying specific natural systems, we rather investigate the simplest model for quantum transport that displays the features mentioned above: a disordered trimer with a superradiant and a subradiant state, on which the excitation is initialized, along with an acceptor state (the RC), where the
excitation can be trapped, see Fig.~\ref{trimer}a. We believe this to be an essential step towards understanding the basic mechanism of transfer optimization in more complex networks.}

Here we discuss the optimization of excitation transfer
efficiency when the excitation is initially prepared 
on a subradiant state rather than on a single site, as is usually done.
There are two main reasons for considering transport from a subradiant initial condition, 
related to the presence of noise and disorder:
(i) The thermalization processes, typically at place in natural complexes, 
such as the LHI-RC of purple bacteria, tend to 
populate the lowest-energy levels, which include both superradiant and subradiant states with respect 
to the RC and to the electromagnetic field \cite{prbdisorder,schulten-lhi}.
(ii) In absence of any noise, subradiant states do not contribute to transport efficiency. 
The presence of moderate static disorder hinders super- and sub-radiance, 
causing subradiant states to become significantly open to transport. 
Hence, a moderate disorder increases the transport efficiency of subradiant states.
At the same time, 
a very large disorder ultimately prevents transport, due to Anderson localization. 
Since disorder is only detrimental for superradiant states, 
the overall optimization of transfer
is determined by the behavior of subradiant states.

%Indeed, since disorder can be  only detrimental when starting
%from a superradiant state, we believe that a general understanding of optimal 
%transport can be obtained only through a detailed knowledge
%of the behavior of the subradiant states.
%The role of superradiant states is also important to obtain quantitative predictions (see~\cite{prbdisorder}) 
%and it will be the subject of future investigations.

%In natural systems the role played by dephasing is crucial and it has been 
%widely investigated in literature \cite{enaqt6}. Therefore,
%we first focus on the interplay of
%static disorder and energy landscape (energy detuning between the
%initial state and the RC) on transport efficiency, subsequently discussing the effect of dephasing.  

To pursue our goal, in Section~\ref{sec:trimer} we present our trimer model. In
Fig.~\ref{trimer}b, the trimer is shown in the super-subradiant
basis to stress the following features: in the absence of disorder, while the subradiant state is
decoupled from the RC, the superradiant state has an enhanced coupling $V_{\rm rc}$  to the RC. 
We discuss the maximization of the average (over disorder) transfer
efficiency with respect to both the detuning $\delta E_{\rm rc}$ between initial
subradiant state and RC and the strength $W$ of disorder. 
We have found two different regimes depending on 
$V_{\rm rc}$ and the energy gap $\Delta$ between the subradiant and superradiant states. 
The optimal detuning is achieved when the energy of the RC is at
resonance with the initial state (subradiant) for $V_{\rm rc} \gg
\Delta$, while we have found a less trivial optimal detuning, namely 
$\delta E_{\rm rc} = V_{\rm rc}^2/\Delta$,
in the case $V_{\rm rc} \leq \Delta$.
This can be interpreted as a resonant condition between dressed states. 
Concerning the optimal disorder, we have found that it
  is always of the order of $V_{\rm rc}$, if
 $V_{\rm rc}$ is larger or comparable with $\Delta$, while it is
 smaller when $V_{\rm rc} \ll \Delta$. 

We analyze in Section \ref{sec:dephasing} how our findings are affected by dephasing.
{Since our aim is to analyze in detail a very simplified model, we include in a paradigmatic way the effects of a dephasing environment, adopting the Haken--Strobl model \cite{hakenstrobl}.
Such an effective master equation assumes white-noise fluctuations of the site energies, thereby implying a high-temperature limit. Even if it is possible to consider more refined phononic baths, they are not necessary here due to the simplicity of our model. Moreover, they have been found to give results that are qualitatively similar to those obtained with the Haken--Strobl approach \cite{srfmo,srrc,others}.
}

We show that dynamical noise induces an additional resonance peak when the RC
energy is equal to the  energy of the superradiant state. Indeed, the incoherent path opened by dephasing
favours transport between those states which are directly coupled, making the final hopping between the superradiant state and the RC the key passage of the transport process. 
By contrast, the coherent resonance condition found in the absence of dephasing
takes into account the interference effects which are present in the system as a whole,
emphasizing the role of the initial condition in determining the transport efficiency.

\section{The disordered trimer model}

\label{sec:trimer}
\begin{figure}
%\begin{indented}\item[]
\includegraphics[width=4.2cm]{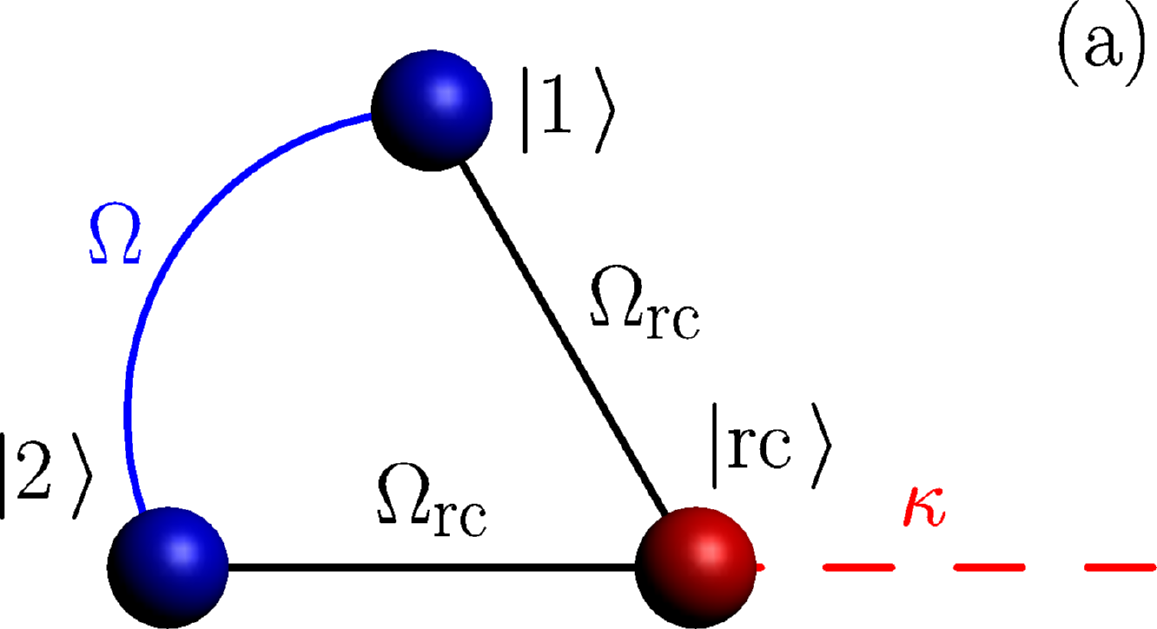}
%\\
\hspace{.0cm}
\includegraphics[width=4.2cm]{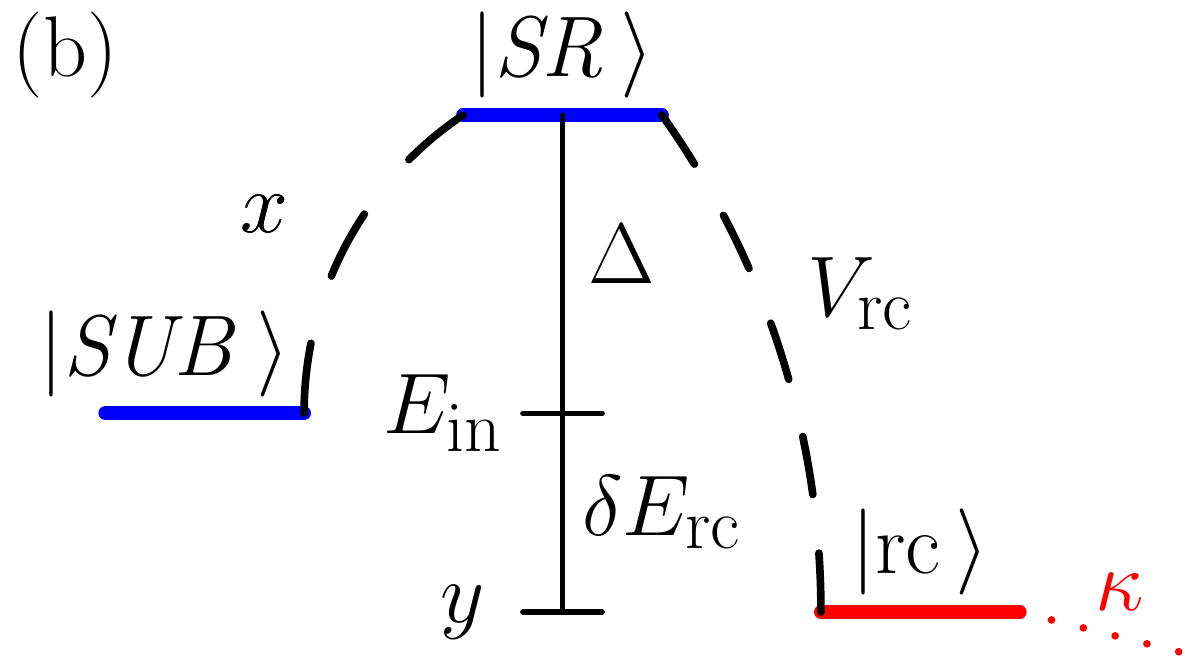}
%\end{indented} 
\caption{(Color online) In panel (a) the trimer model is shown in
  the site basis: two sites connected with the RC
with equal coupling $\Omega_{\rm rc}$ and between them with coupling
$\Omega$. In panel (b) the trimer model is shown in the
superradiant-subradiant-RC basis, see text.
}
\label{trimer}
\end{figure}

The Hamiltonian (in site-basis) for the  trimer model depicted in Fig.~\ref{trimer}a
 can be written in matrix form as follows:
\begin{equation}\label{eq:h3}
\begin{pmatrix}
%\pmatrix{
E_1 - i\frac{\Gamma_{\rm fl}}{2}   & \Omega &  \Omega_{\rm rc}\cr
 \Omega & E_2 - i\frac{\Gamma_{\rm fl}}{2}          &  \Omega_{\rm rc}         \cr
\Omega_{\rm rc}   & \Omega_{\rm rc}          &    E_{\rm rc} - i\frac{\Gamma_{\rm fl}+\kappa}{2}   
\end{pmatrix}\,,
\end{equation}
where the action of the environment (static disorder) has been
taken into account by choosing the energy levels
$E_k$, $k=1,2$, as
Gaussian random numbers
with mean zero and variance $W^2$, and no disorder has been added to the RC site (its
effect can be naturally embedded as additional disorder on the sites $1$ and $2$).

The loss of excitation through the RC has been described by the non-Hermitian term $-i\kappa/2$.
 Throughout the whole paper, $\kappa$ is assumed to be small with respect to the other coupling parameters. 
This choice is consistent with the realistic photosynthetic models.
In order to make a close comparison with realistic systems and
following a standard
procedure~\cite{srfmo,srrc,prbdisorder,alberto,prbdephasing}, we also introduced 
the diagonal non-Hermitian terms $- i\Gamma_{\rm fl}/2$, with the fluorescence constant 
$\Gamma_{\rm fl}$ much smaller than any other energy scale 
($\Omega,\Omega_{\rm rc},W,\kappa$),  representing the loss of excitation 
from each site due to recombination.

It is convenient to move from the site-basis to the subradiant-superradiant-RC basis
by defining the states
\[
\ket{\mathit{SUB}} =  \frac{1}{\sqrt{2}} \left( \ket{1} -\ket{2} \right)\,,
\]
\[
\ket{\mathit{SR}} = \frac{1}{\sqrt{2}} \left( \ket{1} +\ket{2} \right)\,,
\]
from which the new Hamiltonian $H$  easily follows (see also Fig.~\ref{trimer}b):
\begin{equation}
H=
\begin{pmatrix}
%\pmatrix{
-\Omega - i\frac{\Gamma_{\rm fl}}{2}  & x  &  0 \cr
& & \cr
x & \Omega - i\Gamma_{\rm fl}/2          &  \sqrt{2} \Omega_{\rm rc}       \cr
& & \cr
0   & \sqrt{2} \Omega_{\rm rc}          &    y- i\frac{\Gamma_{\rm fl}+\kappa}{2}    
\end{pmatrix}\,,
\label{eq:h3a}
\end{equation}
where the two Gaussian random variables $x=(E_1-E_2)/2$ and $y=E_{\rm rc}-(E_1+E_2)/2$  are such that
$
\langle x \rangle = 0$, $\langle y \rangle = E_{\rm rc}$, and $\langle x^2 \rangle = 
\langle y^2 \rangle -E_{\rm rc}^2 = 
W^2/2$.

In this basis, the coupling between the subradiant state and the RC vanishes, whereas the coupling between
the superradiant state and the RC is enhanced and it is given by the
matrix element $V_{\rm rc}=\sqrt{2} \Omega_{\rm rc}$. 
%Note that such a coupling is enhanced by a factor of $\sqrt{2}$ with respect to the coupling of each single site to the RC. This is a typical example of superradiant enhancement which typically scales as the square root of the system size.
The structure of the Hamiltonian, depicted in Fig.~\ref{trimer}b, implies 
that the excitation transfer from the subradiant state to the RC can only be mediated by the superradiant state, through the random coupling $x$.
{In this basis, the trimer model subradiant-superradiant-RC corresponds to a donor-bridge-acceptor (DBA) system without direct coupling between the donor and the acceptor and with a random coupling between donor and bridge and a random acceptor energy.
In spite of the fact that similar systems have been widely studied in literature, the analysis
developed in the present article concerning the optimization of the decay of an excitation, initialized on the donor state and dissipated at the level of the acceptor state, with the stochastic terms considered here, appears to be new, to the best of our knowledge.
More importantly, our analysis lies outside the regime of validity of perturbation theory, which is used within the superexchange mechanism \cite{MayKuhn} to study these systems. Indeed, the superexchange mechanism can be applied only when the donor--bridge coupling is much smaller than the donor--bridge energy detuning (in our case this detuning can be even zero).}

{A most important quantity for such an analysis (motivated by the study of light-harvesting complexes but also relevant in a general context)} is the efficiency, at the time $t$,  of energy transfer from the system into the RC. 
Given an initial state $\ket{\Psi_{\rm in}}$, it is defined as \cite{enaqt3, enaqt4}
\begin{equation}
\label{eq:eta}
\eta_{_{\Psi_{\rm in}}}(t)=\left\langle\kappa
\int_0^t|\bra{\rm RC}e^{-\frac{i}{\hbar}H\tau}\ket{\Psi_{\rm in}}|^2\,d\tau\right\rangle_{W}\,,  
\end{equation}
and it represents the probability of escaping out of the system  up to the time $t$. In the above definition the brackets $\langle\ldots\rangle_{W}$ indicate the average over disorder.

In numerical simulations we always consider the efficiency at a time
$t\gtrsim \hbar/\Gamma_{\rm fl}$. 
Note that the efficiency strongly depends on the time $t$ at which it is computed for $t<\hbar/\Gamma_{\rm fl}$, while it reaches a stable asymptotic value $\eta^{\infty}$ for $t\gtrsim \hbar/\Gamma_{\rm fl}$, thus motivating our choice. As $\Gamma_{\rm fl}\to 0$, the asymptotic value of the efficiency is $\eta^{\infty}=1$ for any choice of parameters. 
According to a common practice in the study of light-harvesting systems, we will measure energies in cm$^{-1}$ and times in ps.

It is clear from \eqref{eq:eta} that the energy transfer efficiency is strongly dependent
 upon the initial state, a feature also studied in~\cite{fassioli}. 
Indeed, if we start from the superradiant state we are in a situation
in which there is an enhanced direct coupling to the RC at zero
disorder.  Thus, one might think
that the best situation occurs when the excitation is on the
superradiant state set at resonance with the energy of the RC. 
In this situation disorder is  only detrimental to transport,
since it tends to destroy the  superradiant
coupling~\cite{mukameldeph,prbdisorder} and it moves the system out of resonance. 
On the other side, the excitation in natural
complexes is usually spread also on subradiant states, due to the
presence of a thermal bath. Since a subradiant state is not directly
coupled to the RC, it is only through
the action of disorder that the excitation can be transferred from the initial state to the RC (i.e.\ for $W=0$, we have $\eta=0$).

We will focus our attention on this non-trivial case (see Fig.~\ref{trimer}b),
in which
an initial excitation is on the subradiant state at energy $E_{\rm in}=-\Omega$, coupled via $x$ to the superradiant state at energy $E_{\rm sr}=\Omega=E_{\rm in}+\Delta$, which is further coupled to the RC with the
tunnelling amplitude $V_{\rm rc}=\sqrt{2}\Omega_{\rm rc}$. 
Our aim is to find the system configuration that maximizes the
average transfer efficiency. Fixing the energy gap $\Delta=2\Omega$ between the
superradiant and the subradiant states and the superradiant-RC
coupling $V_{\rm rc}$, and assuming $\kappa$ and $\Gamma_{\rm fl}$ to be perturbative quantities, we are left with two independent parameters to be tuned to achieve the maximal efficiency: the subradiant-RC detuning $\delta E_{\rm rc}=E_{\rm rc}-E_{\rm in}$ and the strength $W$ of the random coupling $x$.

To pursue our goal we will first analyze a fully deterministic model obtained replacing the stochastic terms $x$ and $y$ in equation~\eqref{eq:h3a} with deterministic parameters $X$ and $E_{\rm rc}$ as follows:
\begin{equation}
H^{\rm det}=
\begin{pmatrix}
%\pmatrix{
E_{\rm in} - i\frac{\Gamma_{\rm fl}}{2}  & X  &  0 \cr
& & \cr
X & E_{\rm in}+\Delta - i\frac{\Gamma_{\rm fl}}{2}          &  V_{\rm rc}       \cr
& & \cr
0   & V_{\rm rc}          &    E_{\rm rc} - i\frac{\Gamma_{\rm fl}+\kappa}{2}    
\end{pmatrix}\,.
\label{eq:h3d}
\end{equation}
%In particular, we will discuss which pair $(\delta E_{\rm rc}^{\rm opt},X_{\rm opt})$ of values of the subradiant-RC detuning and coupling strength produce the maximal efficiency.
%Subsequently, we will investigate how disorder affects those results.

\begin{figure*}
%\begin{indented}\item[]
\adjustbox{trim={.0\width} {.0\height} {.0\width} {.10\height},clip}%
{\includegraphics[width=8.6cm]{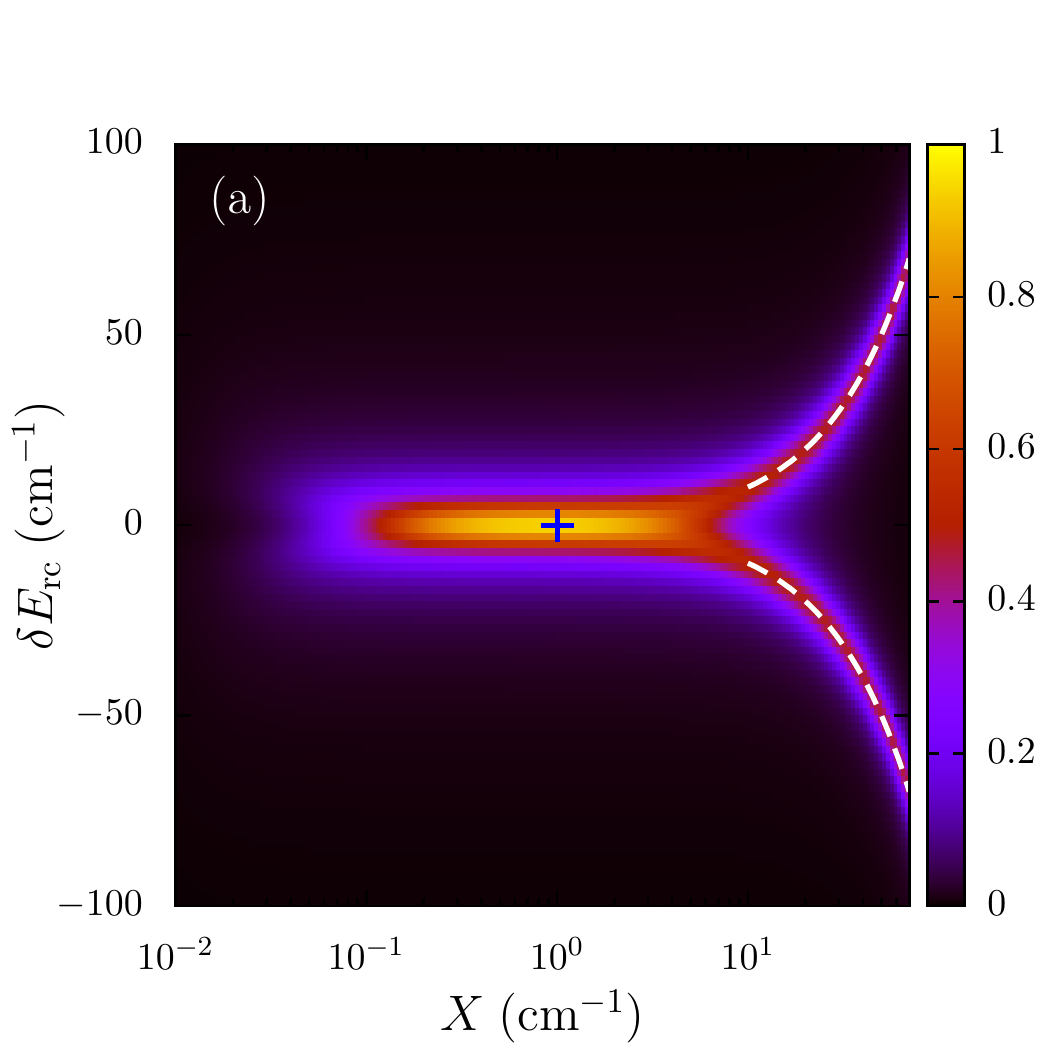}}
\adjustbox{trim={.0\width} {.0\height} {.0\width} {.10\height},clip}%
{\includegraphics[width=8.6cm]{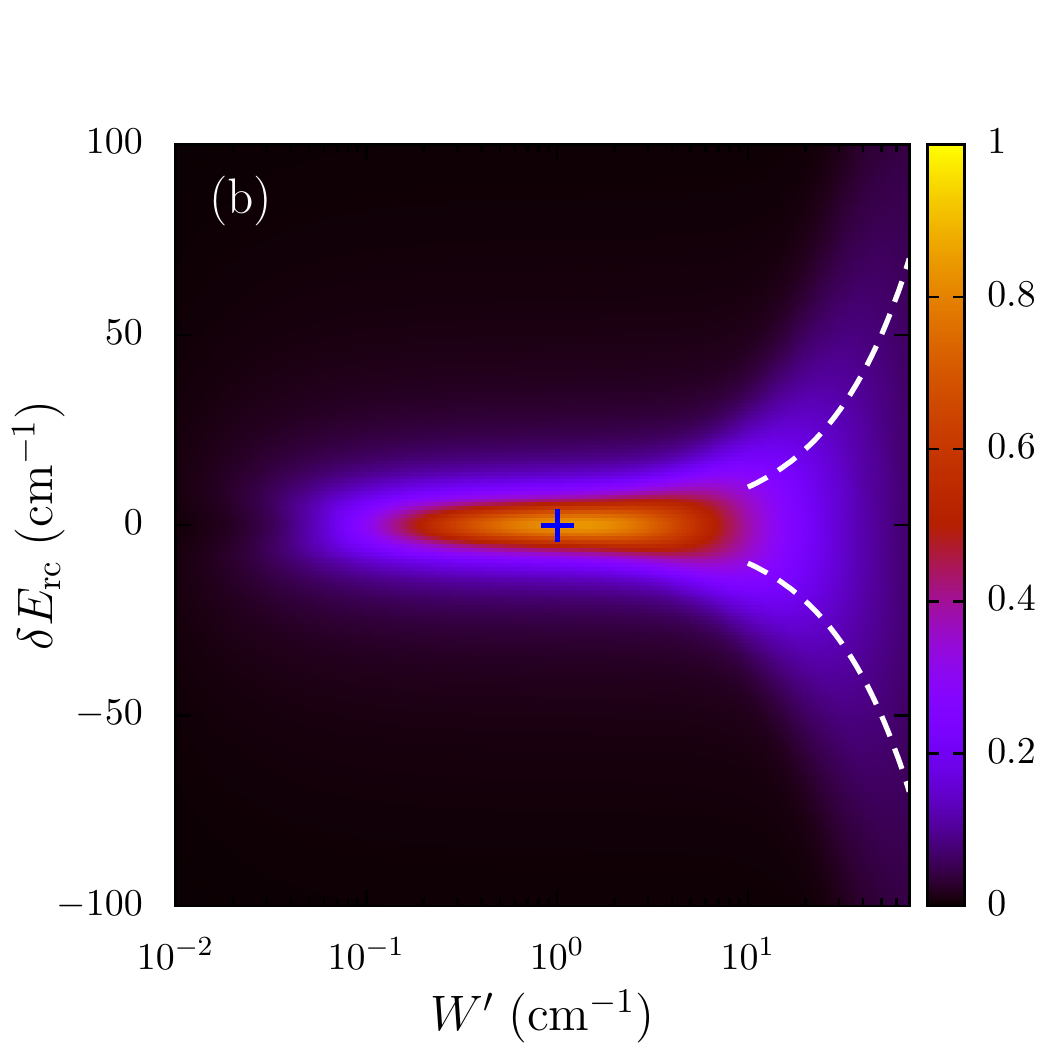}}
%\end{indented} 
\caption{(Color online) Average transfer efficiency $\eta$, computed starting from the subradiant state $\ket{\mathit{SUB}}$, plotted in panel (a) as a function of the subradiant-RC detuning $\delta E_{\rm rc}$ and of the deterministic subradiant-superradiant coupling $X$, and in panel (b) as a function of $\delta E_{\rm rc}$ and of the rescaled disorder strength $W'=W/\sqrt{2}$. The blue cross indicates the estimate~\eqref{eq:gl-opt-d0} for the optimal transfer conditions. The dashed curves indicate the resonances determined in~\eqref{eq:opt-det-xlarge}. The values of the parameters are $\Delta=0\um{cm}^{-1}$, $V_{\rm rc}=1\um{cm}^{-1}$, $\kappa=0.01\um{cm}^{-1}$, and $\Gamma_{\rm fl}=10^{-4}\um{cm}^{-1}$. 
We sampled the efficiency on a $100\times 100$ uniform grid in panel (a) and on a $100\times 200$ uniform grid in panel (b), where the ensemble average over $2000$ realizations of static disorder is shown.
}
\label{fig:delta0}
\end{figure*}

In what follows, we  analyze the results of the deterministic model
comparing them with the results in presence of disorder. In figures \ref{fig:delta0}, \ref{fig:delta1-v10}, and \ref{fig:delta20-v10} the results of the deterministic model are shown in the left panels, while those in presence of disorder, are in the right panels.

\subsection{Optimal disorder and resonance conditions}

According to our previous assumptions, the behavior of the efficiency $\eta$ in both the $(X,\delta E_{\rm rc})$ and $(W,\delta E_{\rm rc})$ planes depends on the ratio between the two system parameters $\Delta$ and $V_{\rm rc}$. We 
first consider the case $\Delta=0$, which corresponds to two uncoupled sites in the trimer model ($\Omega=0$) equally connected to the RC. In this case, the sole energy scale of the system is $V_{\rm rc}$. We will subsequently consider the effects of a finite gap $\Delta\neq 0$.

{We observe that in most of the regimes we consider here, where $\Delta$ is very small compared to other parameters, we cannot rely on the superexchange interaction approach (very effective in other treatment of similar systems where a donor and an acceptor are not directly coupled, see for instance \cite{MayKuhn}) to find optimality conditions, since the perturbative assumptions used there are not generically valid in our context.}

\subsubsection{The zero-gap case: $\Delta=0$.}

With $\Delta=0$ our model corresponds to a tight-binding chain of three sites, 
the first (subradiant) is coupled via $X$ to the second (superradiant), 
with equal energy, which is coupled to the RC with strength $V_{\rm rc}$. It can be easily checked 
that, if we initially excite the subradiant state, 
the probability $P_{\rm RC}(t)$ of finding the excitation on the RC 
(which is a periodic function of time if we neglect the non-Hermitian terms in~\eqref{eq:h3d}) 
can reach the maximal value of $1$ in the shortest time when 
$X=V_{\rm rc}$ and $\delta E_{\rm rc}=0$, thus identifying the following global optimization condition:
\begin{equation}
\label{eq:gl-opt-d0}
X_{\rm opt}=V_{\rm rc}\qquad\text{and}\qquad \delta E_{\rm rc}^{\rm opt}=0\,.
\end{equation}
The estimate given in~\eqref{eq:gl-opt-d0}, obtained considering the
coherent transfer of excitation between the subradiant and the RC
states, is a very good estimate also of the global optimum of the
transfer efficiency, as shown in Fig.~\ref{fig:delta0} (blue cross
in panel (a)).
We observe that the condition $\delta E_{\rm rc}=0$ is not necessary for the probability of being on the RC to reach $1$, but, in combination with $X=V_{\rm rc}$, it makes such transfer the fastest. 

\begin{figure}
%\begin{indented}\item[]
\includegraphics[width=5.7cm]{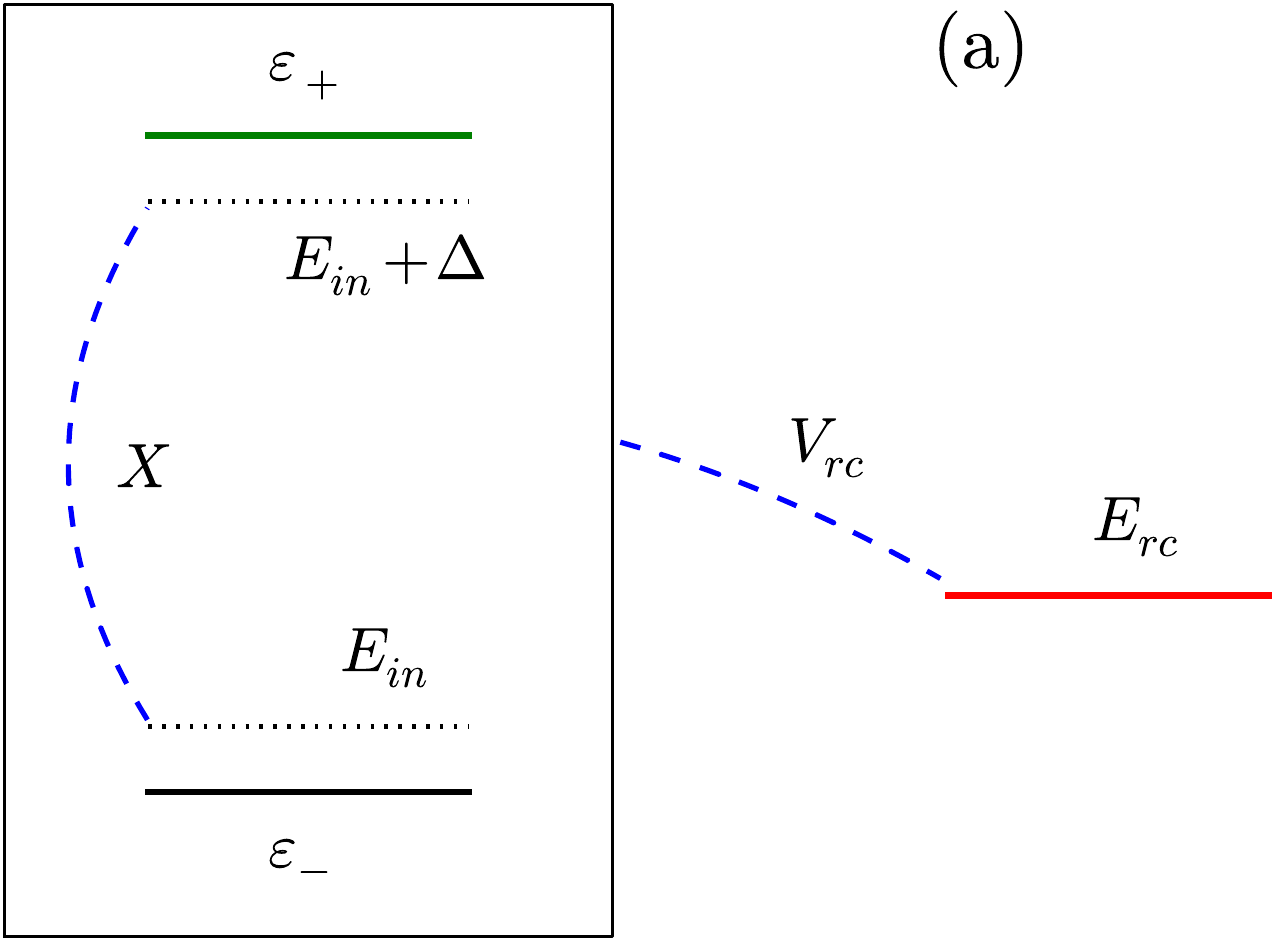}\\
\vspace{.2cm}
\includegraphics[width=5.7cm]{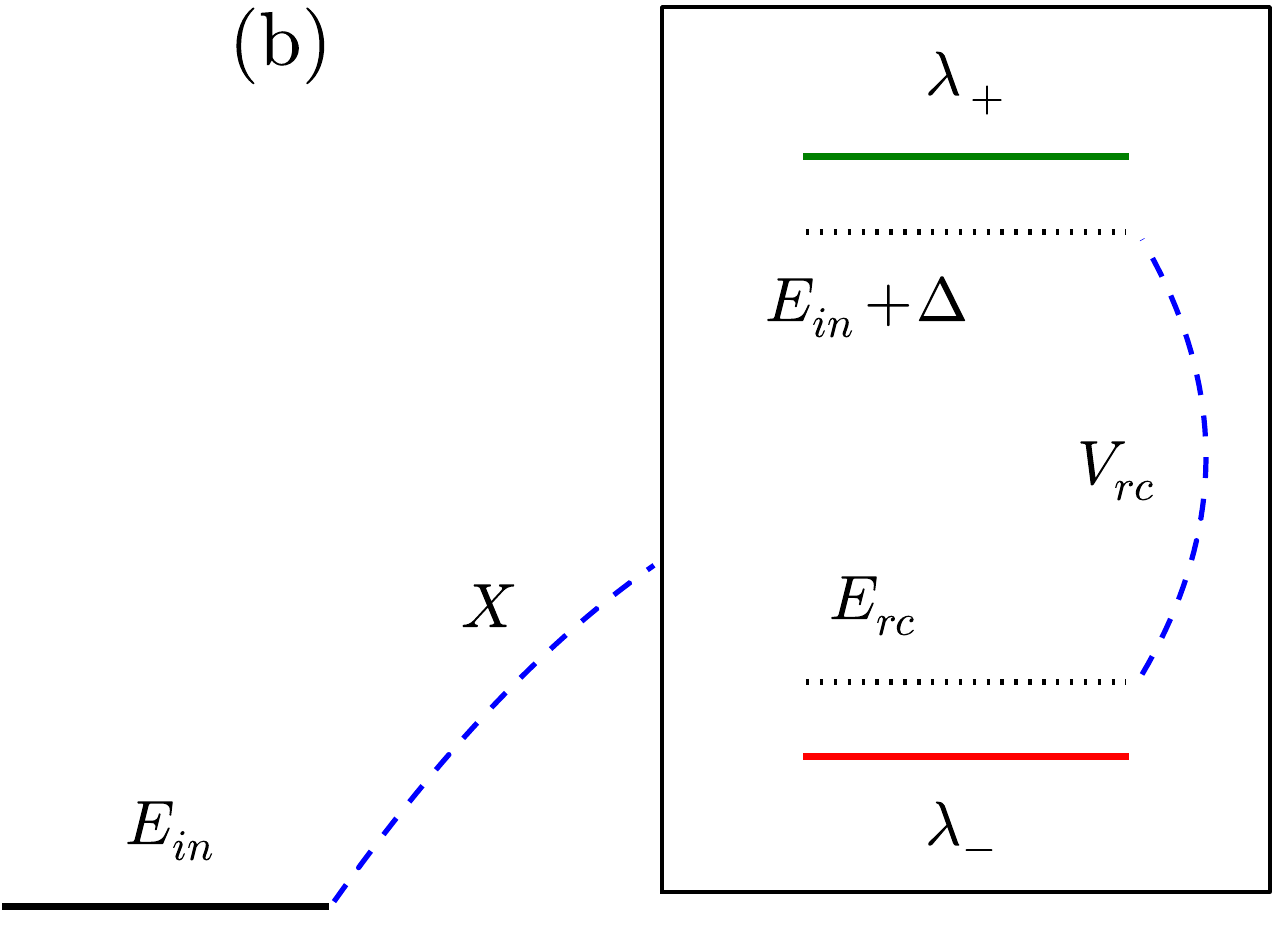}
%\end{indented} 
\caption{(Color online) Different schematic representations of the energy levels in the deterministic trimer model. 
(a) The subradiant-superradiant subsystem (framed), where the coupling $X$ produces the dressed levels with energies $\varepsilon_{\pm}$, is coupled through $V_{\rm rc}$ to the RC state at energy $E_{\rm rc}$.
(b) The initial subradiant state at energy $E_{\rm in}$ is coupled through $X$ to the superradiant-RC subsystem (framed), where the coupling $V_{\rm rc}$ produces the dressed levels with energies $\lambda_{\pm}$.}
\label{fig:2resonances}
\end{figure}

\begin{figure*}
%\begin{indented}\item[]
\adjustbox{trim={.0\width} {.0\height} {.0\width} {.10\height},clip}%
{\includegraphics[width=8.6cm]{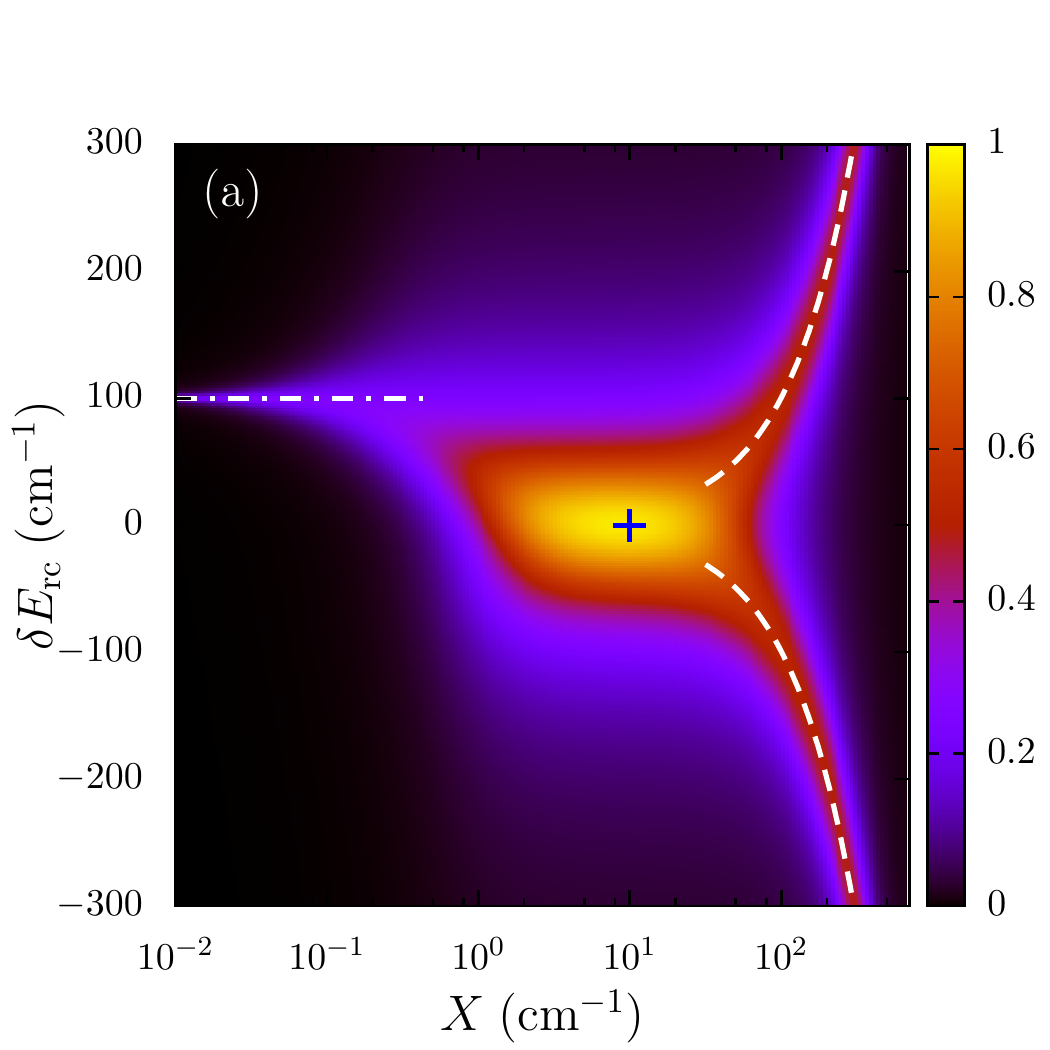}}
\adjustbox{trim={.0\width} {.0\height} {.0\width} {.10\height},clip}%
{\includegraphics[width=8.6cm]{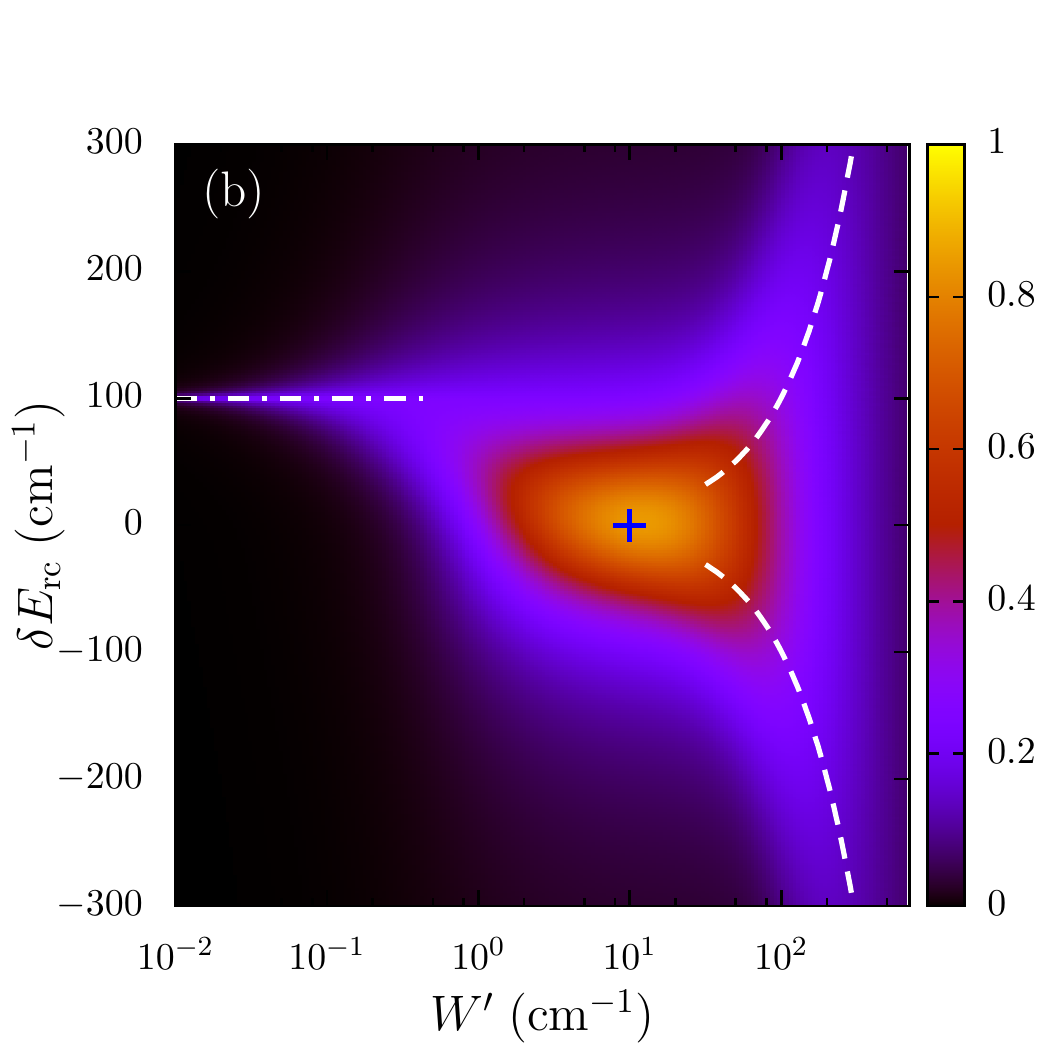}}
%\end{indented} 
\caption{(Color online) Average transfer efficiency $\eta$, computed starting from the subradiant state $\ket{\mathit{SUB}}$, plotted in panel (a) as a function of the subradiant-RC detuning $\delta E_{\rm rc}$ and of the deterministic subradiant-superradiant coupling $X$, and in panel (b) as a function of $\delta E_{\rm rc}$ and of the rescaled disorder strength $W'=W/\sqrt{2}$. The blue cross indicates the estimate~\eqref{eq:gl-opt-d0} for the optimal transfer conditions. The dashed curves indicate the resonances determined in~\eqref{eq:opt-det-xlarge} and the dot-dashed line marks the the resonant condition~\eqref{eq:delta-Eres}. The values of the parameters are $\Delta=1\um{cm}^{-1}$, $V_{\rm rc}=10\um{cm}^{-1}$, $\kappa=0.01\um{cm}^{-1}$, and $\Gamma_{\rm fl}=10^{-4}\um{cm}^{-1}$. We sampled the efficiency on a $100\times 300$ uniform grid in both panels. In panel (b) the ensemble average over $2000$ realizations of static disorder is shown.
}
\label{fig:delta1-v10}
\end{figure*}

On the other hand, if we have some constraint on the coupling $X$ enforcing the condition $X\gg V_{\rm rc}$, the optimal detuning is not given by $\delta E_{\rm rc}=0$. 
To find the  detuning producing the optimal transfer in the case $X\gg V_{\rm rc}$ we can consider $V_{\rm rc}$ as a perturbation, obtaining the picture illustrated in panel (a) of Fig.~\ref{fig:2resonances}. The subradiant and superradiant states couple and give rise to the dressed energy levels
\begin{equation}
\varepsilon_{\pm}(X)=E_{\rm in}+\frac{\Delta}{2}\pm
\sqrt{\frac{\Delta^2}{4} +X^2}\,,
\label{epsilon}
\end{equation}
which reduce to $\varepsilon_{\pm}(X)=E_{\rm in}\pm X$ for $\Delta=0$ (recall that we considered both $\kappa$ and $\Gamma_{\rm fl}$ as small perturbations, that can be neglected in finding the dressed energies). The initial excitation is equally distributed on those levels, and we can then identify two optimal detuning values by the symmetric resonant tunneling conditions
\begin{equation}
\label{eq:res-tunn-xlarge}
E_{\rm rc}=\varepsilon_{\pm}(X)\,,
\end{equation}
entailing
\begin{equation}
\label{eq:opt-det-xlarge}
\delta E_{\rm rc}(X)=\frac{\Delta}{2}\pm
\sqrt{\frac{\Delta^2}{4} +X^2}\,.
\end{equation}
In our case, since $\Delta=0$, we have $\delta E_{\rm rc}(X)=\pm X$ (see dashed curves in Fig.~\ref{fig:delta0}). Note that in~\eqref{eq:opt-det-xlarge} we kept the dependence on $\Delta$, since it will be relevant to what follows.

As can be clearly seen from Fig.~\ref{fig:delta0}, the
results obtained from the  deterministic model are in good agreement
with those in presence of disorder.  Indeed, we can obtain an excellent estimate for the global
optimization condition by simply substituting the deterministic
coupling $X$ with $W/\sqrt{2}$ in \eqref{eq:gl-opt-d0} (blue cross panel (b) of 
Fig.~\ref{fig:delta0}).

\subsubsection{The finite-gap case I: $V_{rc} \gg \Delta$.}\label{sec:finite-gap1}

We now investigate whether the optimal conditions given in
equation~\eqref{eq:gl-opt-d0} are valid also for finite values of the
energy gap $\Delta$. 
%We will show that condition~\eqref{eq:gl-opt-d0}
%gives again a good estimate for the global optimum in the deterministic model, while
%disorder produces some modifications, as it will be discussed in the next subsection.

Let us first consider the situation
where $\Delta<V_{\rm rc}$. We see from Fig.~\ref{fig:delta1-v10} that the global optimization condition~\eqref{eq:gl-opt-d0} (blue cross) is still an excellent estimate for the configuration with maximal efficiency, and also the symmetric resonances present for $X\gg V_{\rm rc}$ follow the analytic prediction~\eqref{eq:opt-det-xlarge} (see dashed curves in Fig.~\ref{fig:delta1-v10}).
Nevertheless, Fig.~\ref{fig:delta1-v10} enlightens a somewhat unexpected feature: if we assume now the coupling $X$ to be constrained within the region $X\ll\Delta<V_{\rm rc}$, the RC energy producing the maximal efficiency, identified by a sharp resonance, is very far from either $E_{\rm in}$ or $\varepsilon_{\pm}$. 

To understand such a resonance, we can now consider $X$ as a small perturbation, exploiting the picture illustrated in panel (b) of Fig.~\ref{fig:2resonances}.
The superradiant and the RC states couple to give the dressed energy levels
\begin{equation}\label{eq:lambda}
\lambda_{\pm} (E_{\rm rc}) =\frac{E_{\rm in}+\Delta+E_{\rm rc}}{2} \pm
\sqrt{\frac{\left(E_{\rm in}+\Delta-E_{\rm rc}\right)^2}{4} +V_{\rm rc}^2}\,,
\end{equation}
which clearly depend on the RC energy $E_{\rm rc}$.
The initial excitation is all on the subradiant state, since $X$ is small, and a resonant tunneling criterion would now require $E_{\rm in}$ to match the energies $\lambda_{\pm}$ of the superradiant-RC subsystem. Nevertheless, we have $E_{\rm in}<E_{\rm in}+\Delta<\lambda_+$ by construction, so that the resonant condition for $X\ll\Delta<V_{\rm rc}$ must be
\begin{equation}
\label{eq:res-tunn-xsmall}
E_{\rm in}=\lambda_-(E_{\rm rc})\,,  
\end{equation}
entailing
\begin{equation}
\label{eq:delta-Eres}
\delta E_{\rm rc}=\frac{V_{\rm rc}^2}{\Delta}\,,  
\end{equation}
which matches exactly the numerical results (see dot-dashed
line in Fig.~\ref{fig:delta1-v10}).

\begin{figure*}
%\begin{indented}\item[]
\adjustbox{trim={.0\width} {.0\height} {.0\width} {.10\height},clip}%
{\includegraphics[width=8.6cm]{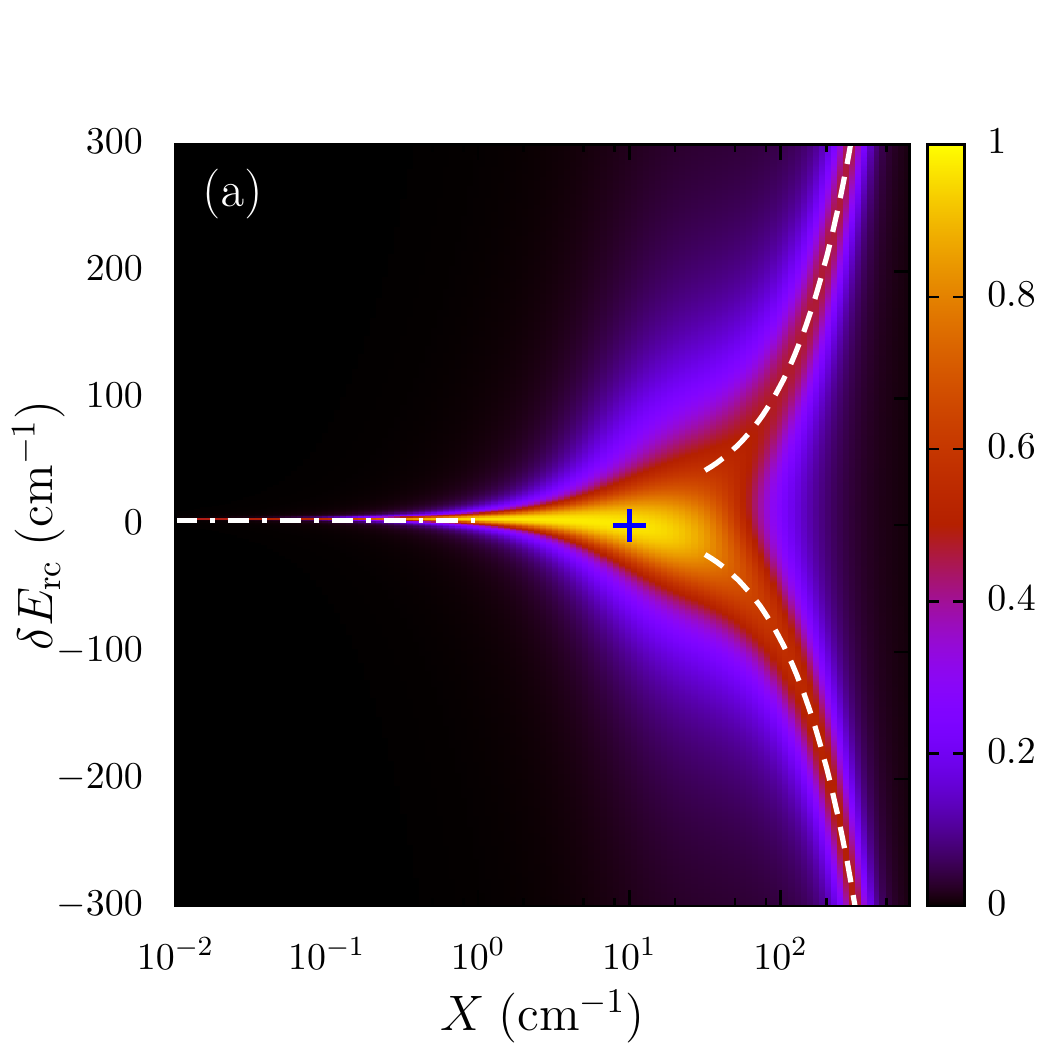}}
\adjustbox{trim={.0\width} {.0\height} {.0\width} {.10\height},clip}%
{\includegraphics[width=8.6cm]{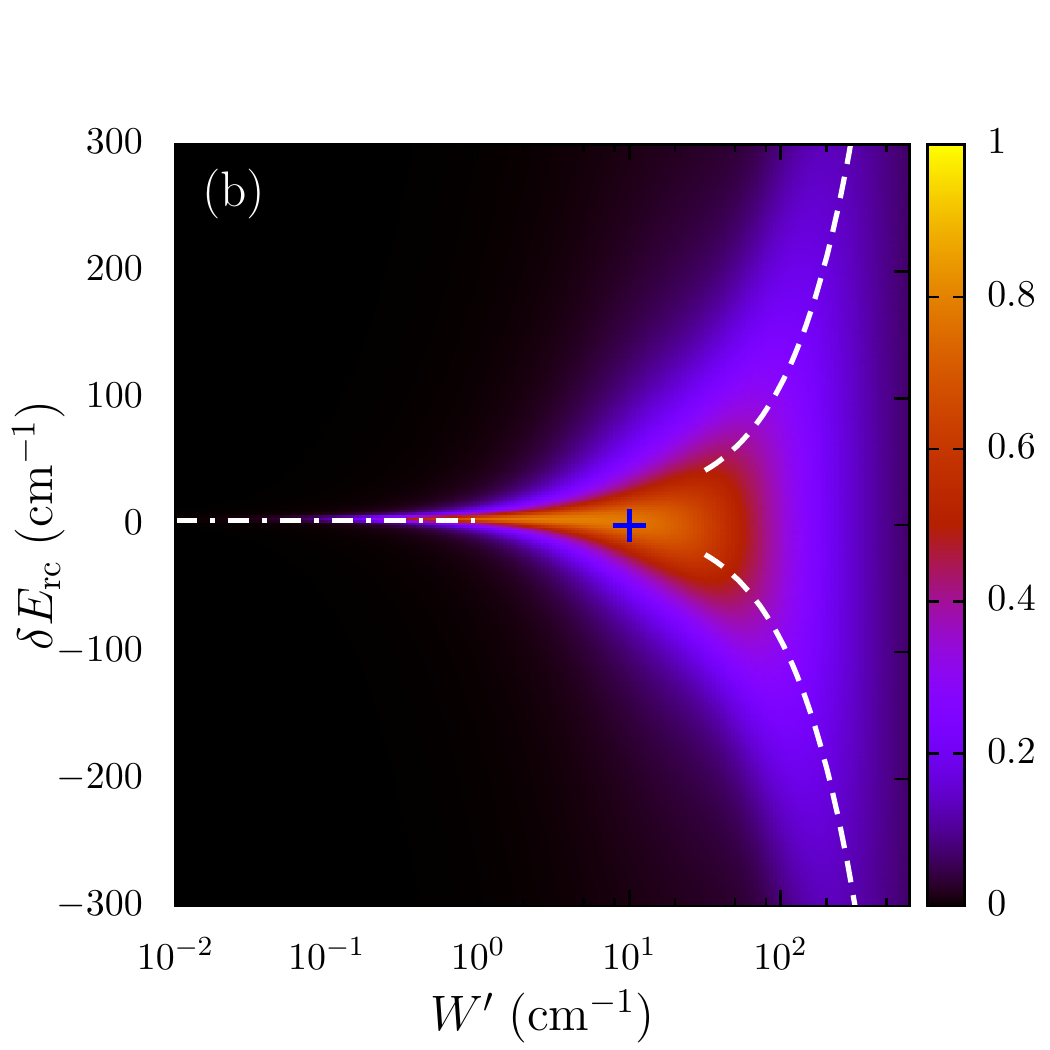}}\\
%\hspace{.1cm}
\adjustbox{trim={.0\width} {.0\height} {.0\width} {.10\height},clip}%
{\includegraphics[width=8.6cm]{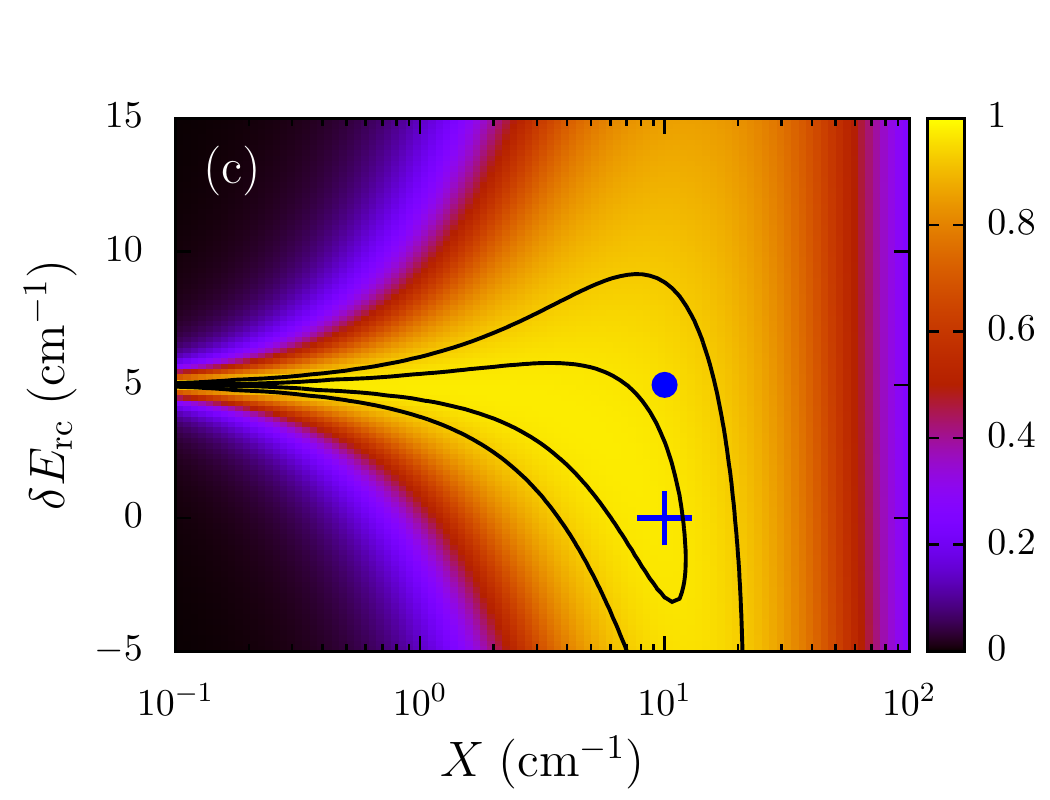}}
\adjustbox{trim={.0\width} {.0\height} {.0\width} {.10\height},clip}%
{\includegraphics[width=8.6cm]{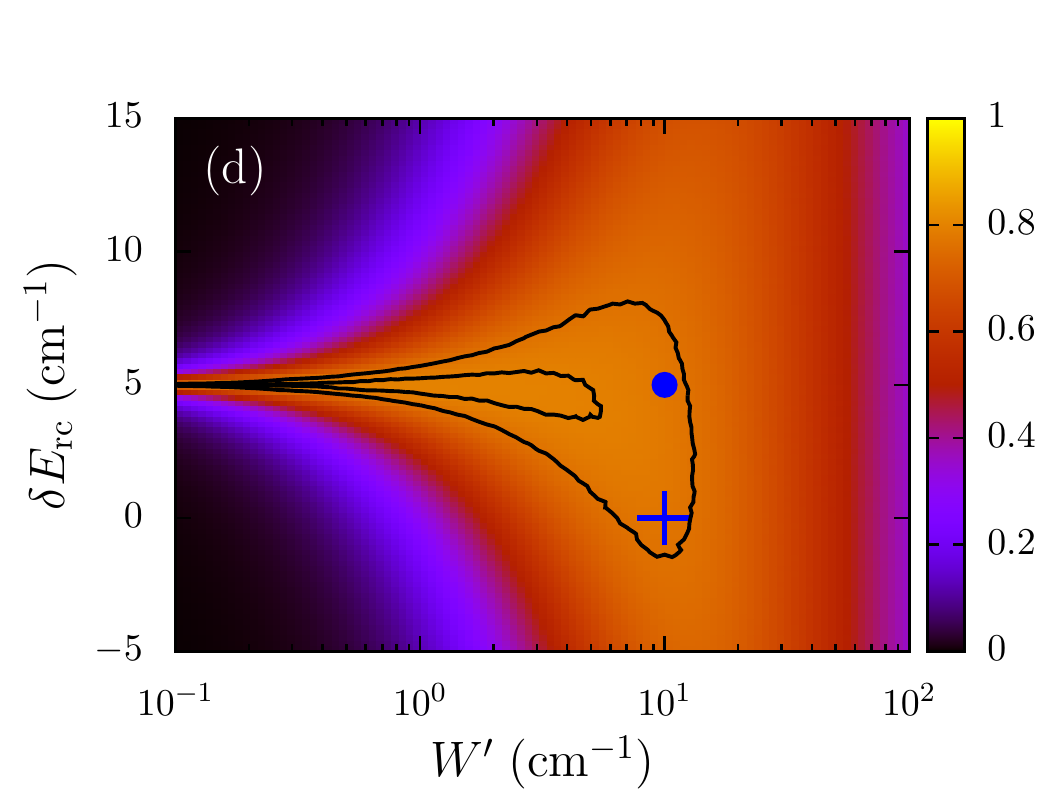}}
%\end{indented} 
\caption{(Color online) Average transfer efficiency $\eta$, computed starting from the subradiant state $\ket{\mathit{SUB}}$, plotted in panels (a) and (c) as a function of the subradiant-RC detuning $\delta E_{\rm rc}$ and of the deterministic subradiant-superradiant coupling $X$, and in panels (b) and (d) as a function of $\delta E_{\rm rc}$ and of the rescaled disorder strength $W'=W/\sqrt{2}$. The blue cross indicates the estimate~\eqref{eq:gl-opt-d0} for the optimal transfer conditions. In panels (a) and (b) the dashed curves indicate the resonances determined in~\eqref{eq:opt-det-xlarge} and the dot-dashed line marks the the resonant condition~\eqref{eq:delta-Eres}. 
In panels (c) and (d) we zoomed on the high-efficiency regions to show how disorder affects the accuracy of the estimates~\eqref{eq:gl-opt-d0} and~\eqref{eq:glob-opt-disorder-l}. The blue dots mark the modification~\eqref{eq:glob-opt-disorder-s} of the estimate proposed in section~\ref{sec:finite-gap}. Isolines enclose the regions of efficiency lying within 1\% and 5\% of the maximal efficiency.
The values of the parameters are $\Delta=20\um{cm}^{-1}$, $V_{\rm rc}=10\um{cm}^{-1}$, $\kappa=0.01\um{cm}^{-1}$, and $\Gamma_{\rm fl}=10^{-4}\um{cm}^{-1}$.
We sampled the efficiency on a $100\times 600$ uniform grid in panels (a) and (b) and on a $60\times 100$ uniform grid in panels (c) and (d). In panels (b) and (d) the ensemble average over $2000$ realizations of static disorder is shown.
}
\label{fig:delta20-v10}
\end{figure*}

%As can be clearly seen from figures
%\ref{fig:delta0} and \ref{fig:delta1-v10}, where $V_{\rm rc}\gg\Delta$, the
%results obtained from the  deterministic model are in good agreement
%with those in presence of disorder.  Indeed, 
Concerning the model in presence of disorder,
we can again obtain an excellent estimate for the global
optimization condition by simply substituting the deterministic
coupling $X$ with $W/\sqrt{2}$ in \eqref{eq:gl-opt-d0} (blue cross
in panel (b) of Fig.~\ref{fig:delta1-v10}).

\subsubsection{The finite-gap case II: $V_{rc} \lesssim \Delta$.}\label{sec:finite-gap}

If we now decrease further the ratio $V_{\rm rc}/\Delta$ we find the
following remarkable result (panel (a) of Fig.~\ref{fig:delta20-v10}): 
the estimate~\eqref{eq:gl-opt-d0}, which was obtained for
$V_{\rm rc} \gg \Delta$, still identifies the global efficiency
optimization in the deterministic model (blue cross). Moreover, the
resonances predicted by~\eqref{eq:opt-det-xlarge} for $X\gg V_{\rm
  rc},\Delta$ and by~\eqref{eq:delta-Eres} for $X\ll V_{\rm
  rc},\Delta$ still correspond to the local optimization of the
efficiency (see white curves in Fig.~\ref{fig:delta20-v10}). As for
the model in presence of disorder (panel (b) of
Fig.~\ref{fig:delta20-v10}), while the estimate~\eqref{eq:gl-opt-d0}
is still within a region of significant efficiency, the optimal
condition is modified by disorder, 
(see panels (c) and (d) of Fig.~\ref{fig:delta20-v10} and the discussion below).

% This robustness of the estimates to the presence of the energy gap $\Delta$ is due to the fact that it only affects the frequency of the oscillations of the probability $P_{\rm rc}(t)$ of finding the excitation on the RC, without preventing it from reaching the value $1$ when $X=V_{\rm rc}$ and 
% $\delta E_{\rm rc}=0$ \textcolor{red}{The whole sentence is not clear}.

Indeed, when $V_{\rm rc}\lesssim\Delta$ (Figure \ref{fig:delta20-v10}), disorder induces some modification of the 
global optimization conditions. This effect can be clearly seen by
comparing panels (c) and (d) of Fig.~\ref{fig:delta20-v10}, which
describe a situation with $V_{\rm rc}/\Delta\approx 0.5$.
The average over disorder shifts the optimal detuning from $\delta E_{\rm rc}=0$ to the low-disorder resonance $\delta E_{\rm rc}=V_{\rm rc}^2/\Delta$ given by~\eqref{eq:delta-Eres}.
This can be explained by the fact that the random coupling falls for
many realizations in the region $X<V_{\rm rc}\lesssim\Delta$, where
the resonance is for $\delta E_{\rm rc}=V_{\rm rc}^2/\Delta$ and not
for $\delta E_{\rm rc}=0$.  
As far as optimal disorder is concerned, even if \eqref{eq:gl-opt-d0}
overestimates its actual  value, it still gives an estimate within
$5\%$ of the maximal efficiency (panel (d) of Fig.~\ref{fig:delta20-v10}). 
In general for $V_{\rm rc} \ll \Delta$ we have
verified that the optimal detuning is still given by  $\delta E_{\rm
  rc}=V_{\rm rc}^2/\Delta$, while the optimal disorder shifts towards
zero as $V_{\rm rc}$ decreases. This is again an effect of the
average over disorder realizations, due to the fact that the resonance
condition  $\delta E_{\rm rc}=V_{\rm rc}^2/\Delta$ is valid for small
disorder. 

\subsection{Summary for the case with only static disorder}
Summarizing,
% as long as the opening strength $\kappa$ and the fluorescence
% constant $\Gamma_{\rm fl}$ can be considered small perturbations
% with respect to the relevant system parameters (a condition
% consistent with realistic applications), 
the global optimization of the average transfer efficiency $\eta(t\gtrsim \hbar/\Gamma_{\rm fl})$ 
from the subradiant state of the trimer into the sink placed at the RC is given by
\begin{equation}
\label{eq:glob-opt-disorder-l}
W_{\rm opt}/\sqrt{2}\simeq V_{\rm rc}\quad\text{and}\quad E_{\rm rc}=E_{\rm in}\quad\text{for}\quad V_{\rm rc}\gg\Delta\,,  
\end{equation}
and
\begin{equation}
\label{eq:glob-opt-disorder-s}
W_{\rm opt}/\sqrt{2}\simeq V_{\rm rc}\quad\text{and}\quad E_{\rm rc}=E_{\rm in}+\frac{V_{\rm rc}^2}{\Delta}\quad\text{for}\quad V_{\rm rc}\lesssim\Delta\,.  
\end{equation}

Another physically relevant question concerns the optimal detuning
at some fixed disorder strength determined by physiological 
conditions (natural systems
are usually subject to a definite range of static disorder).
In this situation our results indicate that, 
if the disorder is constrained to be smaller than 
the energy gap $\Delta$ and the coupling $V_{\rm rc}$, 
the optimal subradiant-RC detuning is not zero, but it is always 
given by
\begin{equation}
\label{eq:delta-eres2}
\delta E_{\rm rc}=\frac{V_{\rm rc}^2}{\Delta}\,,
\end{equation}
with a significant efficiency present only in a narrow
 band around such optimal detuning.
This result is at variance with the intuitive expectation that
the best transport would be obtained 
at resonance with the initial state, $\delta E_{\rm rc}=0$.

For large values of the static disorder, as a result of the averaging procedure, 
we have a broad resonance centered around $\delta E_{\rm rc}=0$, which fades into two 
very broad resonances centered around $E_{\rm rc}=\varepsilon_\pm(W/\sqrt{2})$ given
 by condition~\eqref{eq:res-tunn-xlarge}, characterized by a negligible efficiency. 

\section{The effect of dephasing noise}\label{sec:dephasing}

We now discuss how the presence of dynamical noise affects the coherent features analyzed in the previous section. 
The coupling to a dephasing environment is modeled 
as stochastic fluctuations of diagonal energies, by adding to the
Hamiltonian of the system the time-dependent term
$$
H_{\rm deph}=\hbar \sum_{k=1}^3 q_k(t) |k\rangle \langle k|\,,
$$ 
where the frequencies $q_k(t)$ represent  uncorrelated fluctuations
characterized by
\begin{equation}
\langle q_k(t)q_{k'}(t')\rangle=\frac{\gamma_\phi}{\hbar}\delta_{kk'}\delta(t-t')\,,
\end{equation}
with $\gamma_\phi/\hbar$ being the dephasing rate.
For the sake of comparison with the other system parameters, we represent the intensity of dephasing by the energy parameter $\gamma_\phi$.

This is a common way to include dephasing in exciton 
dynamics, and it gives rise to the Haken--Strobl master equation \cite{hakenstrobl}. 
The Haken--Strobl approach has been widely used in the past to include 
dephasing \cite{mukameldeph,leegwater} and it has also been analyzed in many recent 
applications \cite{prbdephasing, enaqt6, enaqt3, enaqt4, enaqt9, enaqt1, enaqt2, enaqt5, enaqt7, enaqt8, plenioPTA, caoprl} for its simplicity and effectiveness in describing strong dephasing in the high-temperature limit.

{Within this approach, it is possible to analytically perform the average over white noise and 
then consider the evolution of our trimer system as dictated by the 
following master equation for the average density matrix $\rho$ ($h,k=1,2,3$ with $3=\mathrm{RC}$):
\begin{equation}\label{eq:master}
\frac{d\rho_{hk}}{dt}=-\frac{i}{\hbar}(H\rho-\rho H^\dagger)_{hk}-\frac{\gamma_{\phi}}{\hbar}(1-\delta_{hk})\rho_{hk}\,,
\end{equation}
where $H$ is the non-Hermitian trimer Hamiltonian introduced in \eqref{eq:h3}, which contains the time-independent random energies describing static disorder.}

{From the practical point of view, due to the low dimensionality of our model system, we can adopt a very effective computational strategy.
Since Eq.~\eqref{eq:master} is a first-order linear differential equation for the average density matrix $\rho$ governed by a symmetric super-operator, we can diagonalize the latter and then use the exponential map to obtain, with great accuracy, the time-evolution of $\rho$ averaged over dynamical noise.
We do this for each realization of static disorder, performing eventually the ensemble average. We can then say that the average over dynamical noise is treated analytically, while the average over static disorder is performed numerically.}

The transfer efficiency, which depends upon the initial density matrix $\rho^{\rm in}$, is now given by 
\begin{equation}\label{eq:etarho}
\eta_{\rho^{\rm in}}(t)=\left\langle \kappa\int_0^t \rho_{33}(\tau)\,d\tau\right\rangle_W\,.
\end{equation}
where $\langle ... \rangle_W$ represents the average over static disorder.

As before, we study the efficiency at the fluorescence time $t_{\rm fl}=\hbar/\Gamma_{\rm fl}$. It is indeed clear that this is an important time-scale for the system, since most of the excitation goes away within $t_{\rm fl}$. Consequently, if the dephasing $\gamma_\phi$ is smaller than the fluorescence width $\Gamma_{\rm fl}$, it will not have enough time to significantly affect the system dynamics. On the other hand, when $\gamma_\phi>\Gamma_{\rm fl}$, the dephasing will affect the various coherent features of the dynamics within the trimer, as we discuss below.

\subsection{The zero-gap case} 

\begin{figure}
%\begin{indented}\item[]
\includegraphics[width=8.6cm]{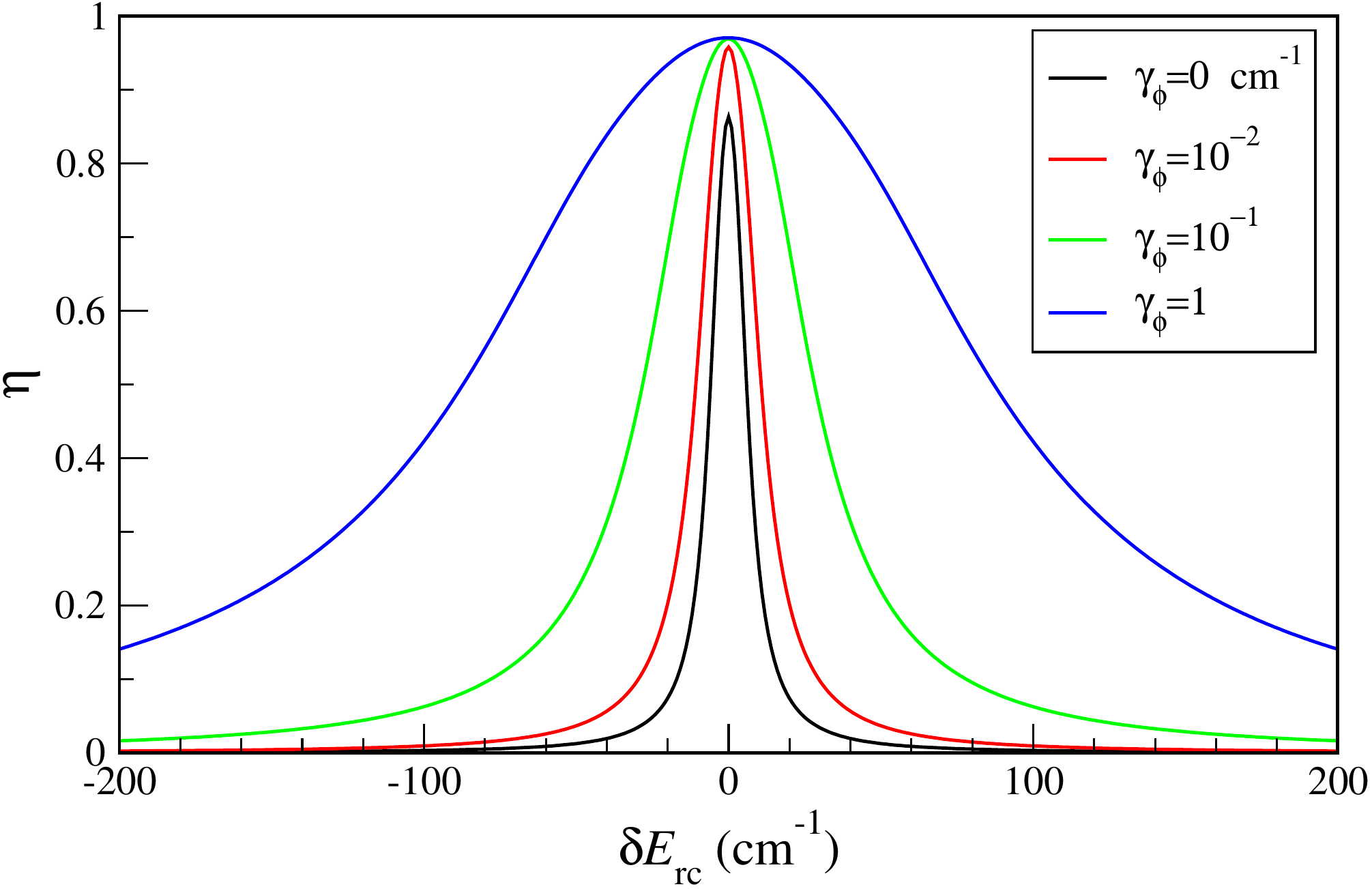}
%\end{indented} 
\caption{(Color online) Average transfer efficiency $\eta$, 
computed starting from the subradiant state 
$\ket{\mathit{SUB}}$ for the optimal disorder
(in absence of dephasing noise) $W_{\rm opt}=\sqrt{2}\um{cm}^{-1}$, 
plotted as a function of the subradiant-RC 
detuning $\delta E_{\rm rc}$ (uniform grid with $100$ points) for different values of the dephasing strength $\gamma_\phi$,
as indicated in the legend.
The values of the parameters are $\Delta=0\um{cm}^{-1}$, $V_{\rm rc}=1\um{cm}^{-1}$, $\kappa=0.01\um{cm}^{-1}$, and $\Gamma_{\rm fl}=10^{-4}\um{cm}^{-1}$, the same as in Fig.~\ref{fig:delta0}. The ensemble average is over $2000$ realizations of disorder.
}
\label{fig:deph2}
\end{figure}

When considering the case with $\Delta=0$, simple symmetry 
considerations show that the optimal detuning must remain 
$\delta E_{\rm rc}^{\rm opt}=0$ also in the presence of dephasing. 
In order to confirm this, we fix  $W=W_{\rm opt}$ (see
\eqref{eq:glob-opt-disorder-l}) and study the efficiency as a
function of the detuning 
$\delta E_{\rm rc}$ for different values
of the dephasing strength $\gamma_{\phi}$.

The results are reported in Fig.~\ref{fig:deph2}.
First of all, we  observe that the presence of dephasing
 makes the system more and more resistant to detuning 
between the initial state and the RC. Indeed, 
while the efficiency is still maximized for $\delta E_{\rm rc}=0$, 
 it remains high in a region which grows with $\gamma_\phi$.
This is consistent with the fact that, while the width
of the resonant peak in the variable $\delta E_{\rm rc}$ for $\gamma_{\phi}=0$
 can be considered as an effect
of quantum coherence, the incoherent dynamics generated by
the Haken--Strobl master equation gives rise to higher 
efficiency even off-resonance.

\begin{figure}
%\begin{indented}\item[]
\includegraphics[width=8.6cm]{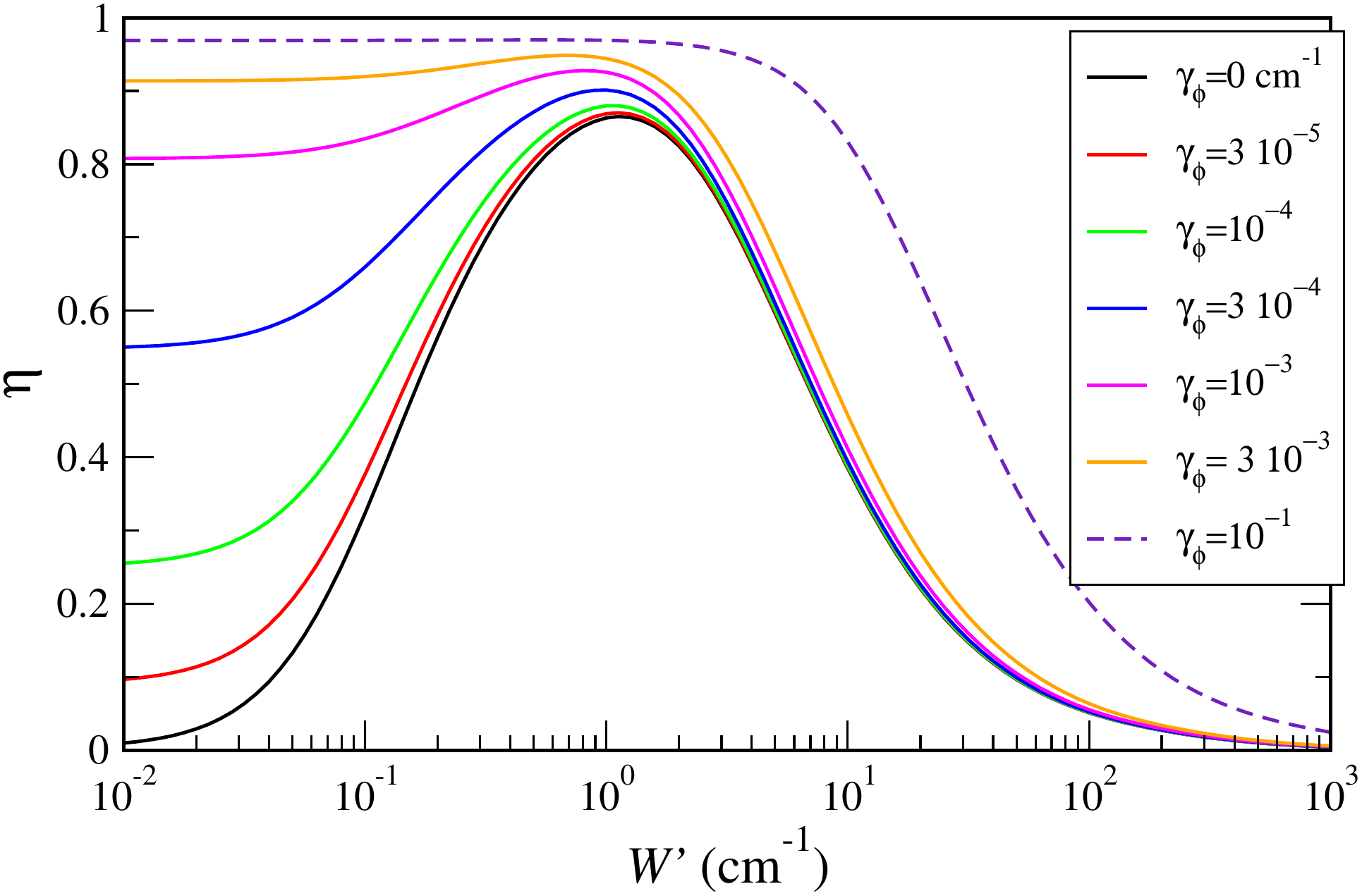}
%\end{indented} 
\caption{(Color online) Average transfer efficiency $\eta$, computed starting 
from the subradiant state $\ket{\mathit{SUB}}$ at resonance with the 
RC ($\delta E_{\rm rc}=0$, which is the optimal detuning
in absence of dephasing)
plotted as a function of the rescaled 
disorder strength $W'=W/\sqrt{2}$  (uniform grid with $100$ points) for different values 
of the dephasing strength (see legend).
The values of the parameters are 
$\Delta=0\um{cm}^{-1}$, $V_{\rm rc}=1\um{cm}^{-1}$, 
$\kappa=0.01\um{cm}^{-1}$, and $\Gamma_{\rm fl}=10^{-4}\um{cm}^{-1}$, the same as in Fig.~\ref{fig:delta0}.
The ensemble average is over $2000$ realizations of disorder.
}
\label{fig:deph1}
\end{figure}

In order to show how the optimal disorder is affected by dephasing, 
we fix $\delta E_{\rm rc}=\delta E_{\rm rc}^{\rm opt}=0$ (optimal
detuning in absence of dephasing)
and plot the correspondent efficiency as a function of the static 
disorder $W'=W/\sqrt{2}$ for different values of the dephasing $\gamma_{\phi}$,
see Fig.~\ref{fig:deph1}.
The dephasing, when not too 
strong compared with the fluorescence width, does not change
the position of the maximal efficiency at $W_{\rm opt}$.

A remarkable effect is that dephasing (if not too strong) enhances transport for 
small disorder strength $W < W_{\rm opt}$, while it is essentially
irrelevant for $W> W_{\rm opt}$. 
This is consistent with the noise-assisted transport picture~\cite{enaqt3,enaqt4},
which predicts that the efficiency is generally enhanced by dephasing,
but it also suggests that such noise-assisted 
transport is more efficient when the energy levels are sufficiently
close to each other. 
For sufficiently large dephasing strength, the efficiency becomes a decreasing function of $W'$, in such condition static disorder is only detrimental to transport 
(see dashed curve in Fig.~\ref{fig:deph1}).

%--------------------------------------------------------------

\subsection{The finite-gap case}

\begin{figure*}
%\begin{indented}\item[]
\adjustbox{trim={.0\width} {.0\height} {.0\width} {.10\height},clip}%
{\includegraphics[width=8.6cm]{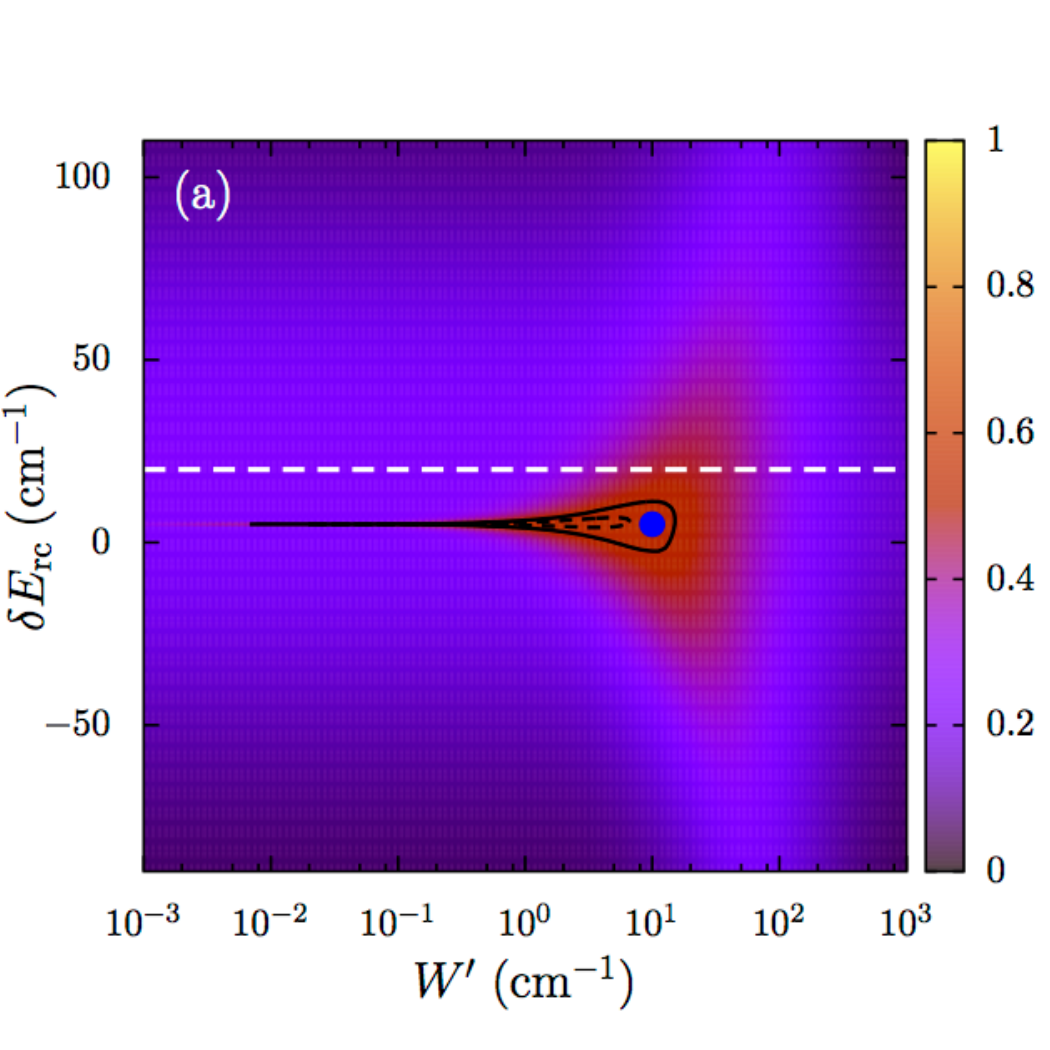}}
\adjustbox{trim={.0\width} {.0\height} {.0\width} {.10\height},clip}%
{\includegraphics[width=8.6cm]{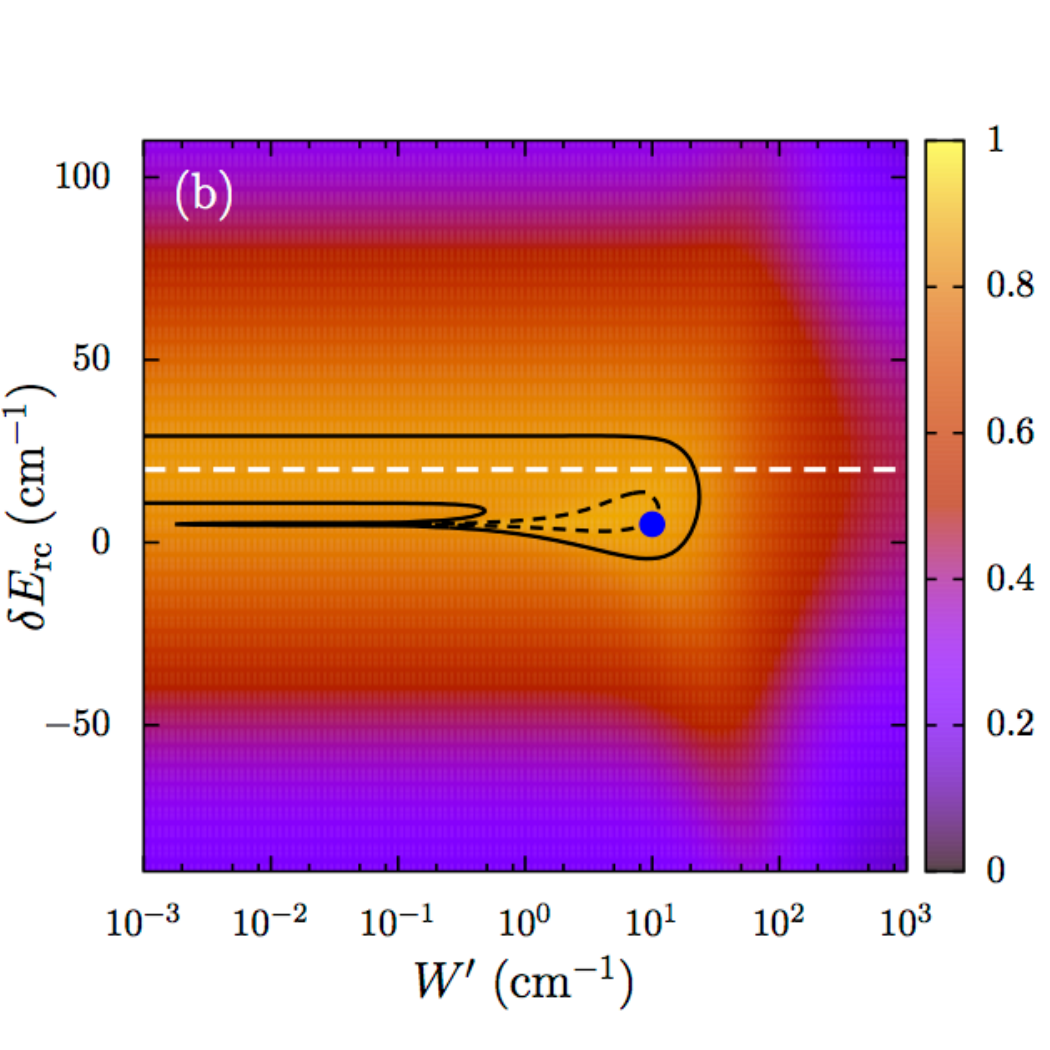}} 
%\end{indented} 
\caption{(Color online) Average transfer efficiency $\eta$, computed
  starting from the subradiant state $\ket{\mathit{SUB}}$, plotted
  as a function of the subradiant-RC detuning $\delta
  E_{\rm rc}$ and of the rescaled disorder strength $W'=W/\sqrt{2}$. 
In panel (a) the case of small dephasing $\gamma_{\phi}=10^{-4}\um{cm}^{-1}$ 
is shown and it should be compared with the results
shown in Fig.~\ref{fig:delta20-v10}. In panel (b) the case of
larger dephasing $\gamma_{\phi}=1.26\, 10^{-3}\um{cm}^{-1}$ is shown. 
The blue dot indicates the estimate~\eqref{eq:glob-opt-disorder-s} for the
optimal transfer conditions, while the horizontal dashed white line
represents the condition $\delta E_{\rm rc}=\Delta$. 
Isolines enclose the regions of efficiency lying within 1\% (dashed
curve) and 5\% (full curve) of the maximal efficiency.
The values of the parameters are $\Delta=20\um{cm}^{-1}$, $V_{\rm rc}=10\um{cm}^{-1}$, $\kappa=0.01\um{cm}^{-1}$, and $\Gamma_{\rm fl}=10^{-4}\um{cm}^{-1}$, the same as in Fig.~\ref{fig:delta20-v10}.
We sampled the efficiency on a $120\times 400$ uniform grid. The ensemble average is over $2000$ realizations of static disorder.
}
\label{fig:delta20-v10b}
\end{figure*}

The results of the previous section show that, for the zero-gap case, the presence of dephasing does not affect the estimate for the optimal detuning and also the optimal disorder is modified only for a quite large dephasing strength.

On the contrary, in the presence of a non-vanishing gap 
$\Delta$ between the subradiant initial state and the superradiant
one, dephasing changes the situation in a remarkable way. 
In Fig.~\ref{fig:delta20-v10b} the average efficiency in the
plane $(W',\delta E_{\rm rc})$ is shown for two different values of the dephasing
strength. The system parameters used in this figure coincide with those of Fig.~\ref{fig:delta20-v10}, except for the dephasing. 
 In panel (a) we present the case of small dephasing
$\gamma_{\phi}=10^{-4}\um{cm}^{-1}$: the optimal conditions are
very close to the case of no dephasing (Fig.~\ref{fig:delta20-v10}). Indeed, the blue dot, which represents the optimal conditions without dephasing, lies between
the curves enclosing the regions of $1\%$ and $5\%$ below the maximal efficiency, and the small-disorder resonance condition \eqref{eq:glob-opt-disorder-s} is perfectly reproduced.

On the other side, in panel (b) of Fig.~\ref{fig:delta20-v10b}, where the case
of a larger dephasing is shown ($\gamma_{\phi}=1.26\, 10^{-3}\um{ cm}^{-1}$),
the picture is radically different. A second important resonance at $\delta
E_{\rm rc}=\Delta$ appears (indicated by a dashed horizontal white
line). This resonance, absent for small dephasing, corresponds to the RC energy being equal
to the energy of the superradiant state. 
Even if our previous estimate of
optimal disorder still lies between
the curves enclosing the regions of $1\%$ and $5\%$ below the maximal efficiency
(blue circle in panel (b) of Fig.~\ref{fig:delta20-v10b}), the
structure of the contour lines is very different for two
reasons: (i) transport is strongly enhanced by dephasing for small disorder;
 (ii) due to the additional broad resonance at $\delta
E_{\rm rc}=\Delta$, we have two resonance peaks instead of one (compare panels (a) and (b) of Fig.~\ref{fig:delta20-v10b}). 
For a larger dephasing noise, only the broad resonance at $\delta
E_{\rm rc}=\Delta$ remains.

We can thus conclude that the
resonance $\delta E_{\rm rc}=V_{\rm rc}^2/\Delta$, 
given by~\eqref{eq:delta-Eres}, coexists, in the presence 
of small dephasing, with another resonance 
at $\delta E_{\rm rc}=\Delta$, which eventually dominates 
the dynamics for large dephasing. The latter condition shows that, 
even if we start from the subradiant state, 
the incoherent dynamics generated by dephasing rests upon 
the direct transfer to the RC from the superradiant state, 
since the latter is continuously populated by the action 
of dephasing. This is at variance with the 
coherent resonance conditions  found in the previous section.

\section{Conclusions}\label{sec:conclusions}

We considered a paradigmatic model of quantum network, namely a trimer
 in which two  sites are coupled to a third site, representing a reaction center 
where the excitation can be trapped. 
The optimal conditions of energy transfer are analyzed in presence of different kind of disturbances: static disorder and dynamical noise.
In similar networks the states can be classified 
as superradiant or subradiant, based on how they transfer 
the excitation into the reaction center.

Subradiant states are not directly coupled to the
 reaction center, but static on-site
disorder can effectively couple them 
with superradiant states. This opens an indirect path for the 
transfer of excitation
from subradiant states to the reaction center,
 mediated by the superradiant state. 
The static disorder which activates the transfer from subradiant states, 
when too strong, hinders transport, so that an optimal disorder 
condition can be determined.

We analyze in detail such a model.
Four parameters determine the different regimes in which the trimer 
can operate: the energy distance $\Delta$ between the subradiant and 
the superradiant state, the intensity $W$ of static disorder, which
induces a  random coupling 
between subradiant and superradiant states, the direct coupling $V_{\rm rc}$ between 
the superradiant state and the reaction center, and the detuning 
$\delta E_{\rm rc}$ between the subradiant initial state and the reaction center.

We study how to optimize the energy transfer efficiency by varying 
both the disorder strength $W$ and the detuning $\delta E_{\rm rc}$. 
The optimal conditions in absence of dephasing noise are given by:
\begin{itemize}
\item[(i)] $W\simeq V_{\rm rc}$ and $\delta E_{\rm rc}=0$ for $V_{\rm rc}\gg\Delta$; 
\item[(ii)] $W\simeq V_{\rm rc}$ and $\delta E_{\rm rc}={V_{\rm rc}^2}/{\Delta}$ for $V_{\rm rc}\lesssim\Delta$.
\end{itemize}

The origin of conditions (i) and (ii) can be synthetically expressed 
as follows.
The disorder strength $W$ determines the amplitude of the coupling between the
subradiant donor state (D) and the superradiant bridge state (B), while $V_{\rm rc}$ is the 
coupling between the superradiant bridge state (B) and the RC acceptor state (A). 
Thus the condition $W\simeq V_{\rm rc}$ implies a symmetry between the D--B and B--A couplings, which 
optimizes transport.
Moreover, the condition $\delta E_{\rm rc}=0$ corresponds to a donor and an acceptor with the same energy, which also helps transport in a coherent regime.
On the other hand, when $W\simeq V_{\rm rc} \lesssim \Delta$, the stochastic nature of the D--B coupling induces a shift of the optimal detuning, which is now determined by the presence, for small values of the D--B coupling, of a sharp resonance in the transfer efficiency at $\delta E_{\rm rc}={V_{\rm rc}^2}/{\Delta}$.

Regarding the effect of dynamical noise (modeled as pure dephasing) 
on the various coherent features we analyze, we show that 
our condition for the optimal disorder is still valid for a not too strong dephasing. On the other hand, the optimal detuning can be strongly
modified, since dephasing produces an additional resonance peak,
corresponding to the superadiant bridge energy being equal to that of the RC acceptor. Such
peak coexists with the one obtained via the coherent resonance conditions given above.
Dephasing produces a general enhancement of the transfer efficiency 
for small disorder, providing also a significant stability with 
respect to the energy detuning $\delta E_{\rm rc}$. 

We believe that the principles of our analysis, even if it has been carried out on a very simple model, are relevant to all those quantum networks in which a 
subradiant (or trapping-free) subspace is present.
Specifically, we think that the interplay of an incoherent resonant condition with a coherent one 
is a generic feature of those quantum networks aimed at modeling realistic light-harvesting complexes, such as ring structures with a trapping site at their center akin to those observed in natural LHI complexes.

Especially to understand the enhancement in transport 
due to disorder, which is ubiquitous in natural light-harvesting complexes, it is 
necessary to go beyond the standard perturbative analysis, since the coupling between peripheral sites and the reaction center is coherently enhanced by superradiance, producing a nontrivial interplay with the various sources of noise and disorder.

%\ack  %command only for IOP
\begin{acknowledgements}
The authors acknowledge useful discussions with Lev Kaplan, Debora Contreras-Pulido, and
Robin Kaiser. G.G.~gratefully acknowledges the support of the Mathematical Soft Matter Unit of the Okinawa Institute of Science and Technology.
\end{acknowledgements}

%\section*{References} %command only for IOP

\end{document}